\documentclass[11pt]{article}
\usepackage{epsfig} 
\usepackage{amssymb}
\usepackage{float} 
\newcommand{\sect}[1]{ \section{#1} \setcounter{equation}{0} }

\newcommand{\Deltaslash}{\Delta \! \! \! \! / \,}
\newcommand{\half}{\mbox{\small{$\frac{1}{2}$}}}
\newcommand{\third}{\mbox{\small{$\frac{1}{3}$}}} 
\newcommand{\quarter}{\mbox{\small{$\frac{1}{4}$}}} 
\newcommand{\threehalves}{\mbox{\small{$\frac{3}{2}$}}} 
\newcommand{\bare}{\mbox{\footnotesize{o}}}
 
\newcommand{\BZss}{{\mbox{\scriptsize{BZ}}}} 
\newcommand{\MSbar}{\overline{\mbox{MS}}} 
 
\newcommand{\Nc}{N_{\!c}}
\newcommand{\Nf}{N_{\!f}}

\newcommand{\Cc}{\mathbb{C}}

\begin{document}
\title{Four loop renormalization of $3$-quark operators in QCD}

\author{J.A. Gracey, \\ Theoretical Physics Division, \\ 
Department of Mathematical Sciences, \\ University of Liverpool, \\ P.O. Box 
147, \\ Liverpool, \\ L69 3BX, \\ United Kingdom.} 
\date{} 
\maketitle 

\vspace{5cm} 
\noindent 
{\bf Abstract.} We renormalize generalized $3$-quark operators that relate to
baryon states using the method devised by Kr\"{a}nkl and Manashov at four loops
in the $\MSbar$ scheme. The anomalous dimensions of the four core operators 
used to compute nucleon matrix elements are determined and their associated 
critical exponents are studied in the conformal window using the Banks-Zaks 
expansion.

\vspace{-16.2cm}
\hspace{13.4cm}

\newpage

\sect{Introduction.}

It has been long-established that quarks are the fundamental quanta that bind
together to form mesons and baryons. Protons and neutrons for instance are
comprised of up and down quarks with a wealth of other observed particles built
out of the more massive quarks as well as valence gluons. One way to understand
the internal structure and properties of protons, for example, requires 
colliding beams of protons or electrons with other protons. This allows us to
quantify the theory models of the constituent quarks which are encapsulated in 
structure functions or parton distribution functions. The latter are
probability distributions of the valence quarks within, say, a proton which if
known accurately enough allows one to deduce which partons dominate proton
properties such as its overall motion, spin or charge. Moreover having as 
refined a knowledge as possible of structure functions and distributions is 
important for making precise theoretical predictions for strong effect 
background radiation for colliders searching for physics beyond the Standard 
Model. Theoretically structure functions can only be computed with 
nonperturbative field theory methods using Quantum Chromodynamics (QCD) as it 
describes the strong sector. Central to such activities in particular is 
lattice field theory which requires the use of high performance computing 
facilties. Improvements in this infrastructure in recent years has led to 
progress to the extent that operator matrix element evaluation, required for 
structure functions, has been significantly refined. Included in this overall 
vision of the theory connection to experiment is a parallel approach which is 
the use of distribution amplitudes that are also purely nonperturbative 
quantities. These have focused in recent years on the structure of baryons in 
several different set-up configurations. See, for instance, 
\cite{1,2,3,4,5,6,7} for a variety of developments in this area primarily using
lattice field theory. In measuring such matrix elements and amplitudes on the 
lattice, which involve baryon operator Green's functions, the renormalization 
group running of the operators themselves also plays an important role. In 
other words the anomalous dimension of the operators are required with the 
observation that if as much high loop order information as is calculationally 
possible is available then the uncertainties in the extrapolation to the 
continuum as well as the evolution to low energies can be minimized. 

Such continuum high loop order operator renormalization has been carried out 
for many years for quark bilinear operators such as the twist-$2$ Wilson 
operators that describe deep inelastic scattering. These are central to the 
operator product expansion which is the main tool to probe properties of the 
proton via its structure functions. However for the distribution amplitude 
approach the renormalization of baryon operators is a key ingredient for the 
overall research programme. The renormalization of such operators in purely 
$SU(3)$ gauge theories has not received as much attention. For instance, the 
one loop modified minimal subtraction ($\MSbar$) scheme renormalization was 
carried out in \cite{8,9,10} followed by the two loop extension in \cite{11} 
for proton and neutron related operators. In more recent years with interest 
extending beyond these two states a general $3$-quark operator was renormalized
at two loops in \cite{12}. The major benefit of computing the general operator 
anomalous dimension is that the dimensions of baryon states other than the 
proton and neutron can be deduced as a corollary.  Subsequently the work of 
\cite{11,12} was extended to three loops in \cite{13}. With the progress in 
distribution function analyses the higher order perturbative computation moved 
in several directions. To assist with lattice matching not only does the 
programme require the operator anomalous dimensions but the loop expansion of 
pertinent Green's functions is also necessary. For instance, these Green's 
functions involve the insertion of a $3$-quark operator into a quark $3$-point 
function. Such operator matrix element evaluation was partly addressed in 
\cite{13} where the $3$-quark operator was inserted at zero momentum in a quark
$3$-point function with the squared momenta of the three external quark legs 
held at the same value corresponding to a pseudo-symmetric configuration. 
However, in order to probe proton structure further non-zero momentum operator 
insertion configurations are needed and this was addressed perturbatively in 
\cite{14} to two loops. In that work the $3$-quark operator was inserted with a
non-zero external momentum where the momenta of the other three quark legs 
comprising the Green's function of interest were not nullified either. This 
configuration is more accommodating for lattice measurements especially since 
the setup was completely symmetric. By this we mean the square of each of the 
four external momenta were equal. More recently a new phase of $3$-quark 
operator renormalization and Green's function evaluation has opened when the 
first moment of the $3$-quark operators was considered at two loops in 
\cite{7}. So it is clear there is a well-defined programme of research that is 
needed to support and complement lattice matching activities for distribution
amplitudes.

Such perturbative calculations beyond the lowest loop order are not
straightforward. Aside from the large number of Feynman graphs to be evaluated,
in order to determine the relevant $3$-quark operator Green's functions 
requires decomposing them into a basis of Lorentz tensors. For any such Green's
function the basis is very large and each associated form factor needs to be
fully found explicitly perturbatively to be of any benefit for a lattice 
measurement. This article aims to address one aspect of the $3$-quark operator
programme by determining the anomalous dimension of the general operator 
introduced in \cite{12} to four loops in the $\MSbar$ scheme. Consequently we 
will deduce the anomalous dimensions of various baryon states to the same 
order. To achieve this level of loop order we make extensive use of the 
{\sc Forcer} algorithm, \cite{15,16}, written in the symbolic manipulation
language {\sc Form}, \cite{17,18}. The {\sc Forcer} package is the state of the
art tool for extracting divergences from the Green's function configuration we
have chosen for the renormalization of the general $3$-quark operator to four
loops. Since baryon operators have been of recent interest for probing various 
ideas into beyond the Standard Model physics, (see for instance 
\cite{19,20,21,22,23,24,25}), we will use our results to study the behaviour of
the various baryon operator dimensions in the conformal window. This is 
achieved through the Banks-Zaks perturbative approach of expanding around a 
non-trivial zero of the $\beta$-function in this window, \cite{26}. 
Interestingly we were able to re-sum the critical exponents of several 
operators deep into the lower part of the window. By providing benchmark values
of the exponents for various values of $\Nf$, which is the number of quark 
flavours, it would be interesting to see if these tally with values computed by
other techniques such as lattice field theory.

The article is organized as follows. Section $2$ is devoted to the background
of the general $3$-quark operator we will renormalize to four loops including
the computational strategy and the technicalities of the generalized 
$\gamma$ algebra that is employed. The main results of the renormalization are 
recorded in Section 3 where the anomalous dimensions of four core baryon 
operators are deduced from the general $3$-quark operator anomalous dimension. 
The properties of the four baryon operators at the Banks-Zaks critical point 
are discussed in Section 4 before we present concluding remarks in Section 5.
Appendix A records various $\gamma$ matrix identities which were necessary to
translate the operator renormalization constant to the anomalous dimension.

\sect{General operator}

We begin by discussing the renormalization of a general $SU(3)$ gauge invariant
$3$-quark operator which is, \cite{12},
\begin{equation}
{\cal O}^{ijk}_{\alpha\beta\gamma} ~=~ \epsilon^{IJK} \psi^{iI}_\alpha
\psi^{jJ}_\beta \psi^{kK}_\gamma
\end{equation}
where throughout the quarks are massless, $i$, $j$ and $k$ are flavour indices 
and the $SU(3)$ group generators are $T^a_{IJ}$ with $1$~$\leq$~$a$~$\leq$~$8$ 
and $1$~$\leq$~$I$~$\leq$~$3$. At the initial stage this general form for the 
quark content of a baryon operator is renormalized using the same approach as 
\cite{12} since the anomalous dimensions of specific baryon operators are
deduced from an appropriate choice of a $\gamma$ matrix basis. In order to 
exploit the efficient four loop {\sc Forcer} package, \cite{15,16}, written in 
the symbolic manipulation language {\sc Form} \cite{17,18} we will evaluate the
Green's function
\begin{equation}
G_{(\alpha^\prime \beta^\prime \gamma^\prime | 
\alpha \beta \gamma)}^{(i^\prime j^\prime k^\prime | ijk)}(p) ~=~ 
\langle \psi_{\alpha^\prime}^{i^\prime} (p) ~ 
\psi_{\beta^\prime}^{j^\prime} (-p) ~ \psi_{\gamma^\prime}^{k^\prime} (0) ~ 
{\cal O}_{\alpha \beta \gamma}^{ijk}(0) \rangle 
\label{gfdef}
\end{equation}
in dimensional regularization in $d$~$=$~$4$~$-$~$2\epsilon$ dimensions where 
an external momentum flows in one quark leg and out through another. There is 
no momenta flow through the remaining quark leg or the operator itself. The 
setup is illustrated graphically in Figure \ref{figgf}. With quark fields only,
and no anti-quarks, flowing into the operator insertion then closed quark loops
will only occur on internal gluons. An overall flavour tensor can be factored 
off 
$G_{(\alpha^\prime \beta^\prime \gamma^\prime | 
\alpha \beta \gamma)}^{(i^\prime j^\prime k^\prime | ijk)}(p)$ via
\begin{equation}
G_{(\alpha^\prime \beta^\prime \gamma^\prime | 
\alpha \beta \gamma)}^{(i^\prime j^\prime k^\prime | ijk)}(p) ~=~ 
{\cal T}^{(i^\prime j^\prime k^\prime | ijk)}
G_{(\alpha^\prime \beta^\prime \gamma^\prime | \alpha \beta \gamma)}(p)
\end{equation}
where
\begin{equation}
{\cal T}^{(i^\prime j^\prime k^\prime | ijk)} ~=~ 
\delta^{i i^\prime} \delta^{j j^\prime} \delta^{k k^\prime} ~+~ 
\delta^{i i^\prime} \delta^{j k^\prime} \delta^{k j^\prime} ~+~ 
\delta^{i j^\prime} \delta^{j i^\prime} \delta^{k k^\prime} ~+~ 
\delta^{i j^\prime} \delta^{j k^\prime} \delta^{k i^\prime} ~+~ 
\delta^{i k^\prime} \delta^{j i^\prime} \delta^{k j^\prime} ~+~ 
\delta^{i k^\prime} \delta^{j j^\prime} \delta^{k i^\prime} ~.
\end{equation}
The structure that remains, $G_{(\alpha^\prime \beta^\prime \gamma^\prime | 
\alpha \beta \gamma)}(p)$, will involve the tensor product of three strings of
$\gamma$ matrices with open spinor indices. Before the evaluation of an 
individual Feynman graph each of the strings will have contracted Lorentz 
indices across the strings as well as contractions with the internal loop 
momenta. Since the computation was for a fixed colour group $SU(3)$ we made
extensive use of the $SU(\Nc)$ identity
\begin{equation}
T^a_{IJ} T^a_{KL} ~=~ \frac{1}{2} \left[ \delta_{IL} \delta_{KJ} ~-~ 
\frac{1}{\Nc} \delta_{IJ} \delta_{KL} \right]
\end{equation}
for $\Nc$~$=$~$3$ after using the Lie algebra
\begin{equation}
[ T^a, T^b ] ~=~ i f^{abc} T^c
\end{equation}
to replace any structure functions $f^{abc}$ that arise from the gluon and 
ghost vertex Feynman rules.

{\begin{figure}[ht]
\begin{center}
\includegraphics[width=7.5cm,height=7.5cm]{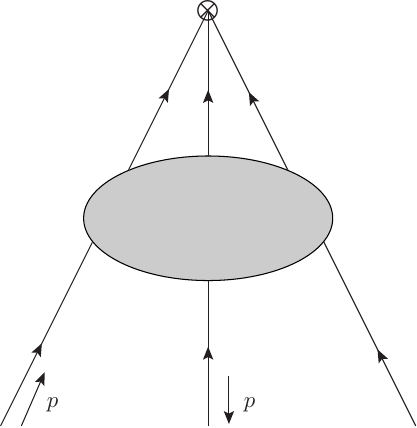}
\end{center}
\caption{Graphical structure of the Green's function 
$G_{(\alpha^\prime \beta^\prime \gamma^\prime | 
\alpha \beta \gamma)}^{(i^\prime j^\prime k^\prime | ijk)}(p)$ defined in
(\ref{gfdef}).}
\label{figgf}
\end{figure}}

One aspect of the {\sc Forcer} package is that it only evaluates Feynman 
integrals which are Lorentz invariants. Therefore loop momenta embedded in 
$\gamma$ matrix strings have to be accommodated first. As in \cite{13} this is 
carried out using what is now a standard projection procedure. It is 
straightforward to withdraw the loop momenta from the $\gamma$ matrices to 
leave a tensor integral for each graph where the rank is an even integer. For 
instance for a four loop graph in the Feynman gauge the highest rank of an 
integral is eight. As each tensor integral has even rank it will be equal to a 
sum of rank eight tensors built from the metric $\eta_{\mu\nu}$ and the sole
external momentum $p_\mu$. There are $764$ independent tensors that the rank 
eight integral decomposes into and mapping the integral to this basis is
achieved by constructing and inverting a $764$~$\times$~$764$ matrix. Clearly 
this is a huge exercise. However as the integral rank is even and there is only
one independent external momentum one can ensure that the matrix is sparse by 
redefining the tensors of the decomposition into transverse, $P_{\mu\nu}(p)$, 
and longitudinal $L_{\mu\nu}(p)$ projectors instead of combinations of
$\eta_{\mu\nu}$ and $p_\mu$ where
\begin{equation}
P_{\mu\nu}(p) ~=~ \eta_{\mu\nu} ~-~ \frac{p_\mu p_\nu}{p^2} ~~~,~~~ 
L_{\mu\nu}(p) ~=~ \frac{p_\mu p_\nu}{p^2} ~.
\end{equation}
The benefit of this choice rests in the algebra of the projectors which is
\begin{equation}
P_\mu^{~\sigma}(p) P_{\sigma\nu}(p) ~=~ P_{\mu\nu}(p) ~~,~~
L_\mu^{~\sigma}(p) L_{\sigma\nu}(p) ~=~ L_{\mu\nu}(p) ~~,~~
P_\mu^{~\sigma}(p) L_{\sigma\nu}(p) ~=~ 0 
\end{equation}
and has the useful property that the matrix is block diagonal. Using this 
approach we have constructed the four general projection tensors up to rank 
eight to carry out the four loop renormalization of the baryon operator in the 
Feynman gauge. As a check on this routine we have reproduced the two and three 
loop results of \cite{12,13} for an arbitrary gauge parameter as originally 
that renormalization was carried out in the Feynman gauge. Therefore all the 
tensor integrals can be written in terms of numerator scalar products and 
$P_{\mu\nu}(p)$ and $L_{\mu\nu}(p)$. The contraction of the Lorentz indices of 
these two projectors with the original $\gamma$ strings produces strings 
involving contracted Lorentz indices or the external momenta.

While the resulting scalar integrals involve scalar products of the external
and loop momenta, which could in principle be passed to the {\sc Forcer}
algorithm in our automated Feynman integral evaluation code, there is a large
amount of associated accompanying $\gamma$ algebra which would slow the actual
integration. As the computation is dimensionally regularized the Clifford
algebra of the four dimensional $\gamma$ algebra is no longer finite 
dimensional but instead is infinite dimensional. Therefore the three open 
$\gamma$ strings need to be mapped to the basis of $d$-dimensional 
$\gamma$ matrices in the same way as the two and three loop computations of
\cite{12,13} were. This basis was introduced and discussed in 
\cite{27,28,29,30,31} and consists of the countably infinite set of matrices 
$\Gamma_{(n)}^{\mu_1 \ldots \mu_n}$ where the integer $n$ lies in the range 
$0$~$\leq$~$n$~$<$~$\infty$. These generalized $\gamma$ matrices are totally 
antisymmetric in the Lorentz indices and defined by
\begin{equation}
\Gamma_{(n)}^{\mu_1\ldots\mu_n} ~=~ \gamma^{[\mu_1} \ldots \gamma^{\mu_n]}
\label{gengamdef}
\end{equation}
where a factor of $1/n!$ is understood. The finer details of the algebra that 
they satisfy is discussed in \cite{30,31}. It is important to remember that the
basis is infinite when the spacetime dimension $d$ is not an integer. If $d$ is
an integer the basis reduces substantially to a finite set. For instance in 
four dimensions it corresponds to
\begin{eqnarray}
\left. \Gamma_{(0)} \right|_{d=4} &=& {\cal I} ~~,~~
\left. \Gamma_{(1)}^{\mu} \right|_{d=4} ~=~ \gamma^\mu ~~,~~
\left. \Gamma_{(2)}^{\mu\nu} \right|_{d=4} ~=~ \sigma^{\mu\nu} \nonumber \\
\left. \Gamma_{(3)}^{\mu\nu\sigma} \right|_{d=4} &=& 
\epsilon^{\mu\nu\sigma\rho} \gamma^5 \gamma_\rho ~~,~~ 
\left. \Gamma_{(4)}^{\mu\nu\sigma\rho} \right|_{d=4} ~=~
\epsilon^{\mu\nu\sigma\rho} \gamma^5  ~~,~~ 
\left. \Gamma_{(n)}^{\mu_1\ldots\mu_n} \right|_{d=4} ~=~ 0 ~~
\mbox{for $n$~$\geq$~$5$} 
\label{gamndto4}
\end{eqnarray}
where ${\cal I}$ is the unit matrix in four dimensions, $\Gamma_{(0)}$ is 
regarded as the $d$-dimensional unit matrix and 
$\sigma^{\mu\nu}$~$=$~$\half [\gamma^\mu, \gamma^\nu]$. The collapse to five 
non-trivial matrices in this dimension means that in 
$d$~$=$~$4$~$-$~$2\epsilon$ dimensions, when $\epsilon$ is small, one can 
regard any operator involving $\Gamma_{(n)}^{\mu_1\ldots\mu_n}$ for 
$n$~$\geq$~$5$ as evanescent. In other words while they are absent in the 
critical dimension of the theory their presence in the regularized theory 
cannot be neglected in the limit to four dimensions when the regularization is 
lifted. This is not unconnected with the definition of $\gamma^5$ which is a 
purely four dimension matrix with no natural analytic continuation to the 
dimensionally regularized theory. However we note that $\gamma^5$ has no 
connection whatsoever with $\Gamma_{(5)}^{\mu_1\dots\mu_5}$. In \cite{32} a 
procedure was developed that allowed for the renormalization of $\gamma^5$ 
dependent quark bilinear operators to high loop order using the $\Gamma_{(n)}$ 
matrices. Briefly this involved the introduction of a finite renormalization 
that restored and ensured the chiral symmetry property of the operators in four 
dimensions was respected.  The generalized $\gamma$ matrices that we employ 
here have several useful properties such as, \cite{29,30,31},
\begin{eqnarray}
\Gamma^{\mu_1 \ldots \mu_n}_{(n)} \gamma^\nu &=&
\Gamma^{\mu_1 \ldots \mu_n \nu}_{(n+1)} ~+~ \sum_{r=1}^n (-1)^{n-r} \,
\eta^{\mu_r \nu} \, \Gamma^{\mu_1 \ldots \mu_{r-1} \mu_{r+1} \ldots
\mu_n}_{(n-1)} \\
\gamma^\nu \Gamma^{\mu_1 \ldots \mu_n}_{(n)} &=&
\Gamma^{\nu \mu_1 \ldots \mu_n}_{(n+1)} ~+~ \sum_{r=1}^n (-1)^{r-1} \,
\eta^{\mu_r \nu} \, \Gamma^{\mu_1 \ldots \mu_{r-1} \mu_{r+1} \ldots
\mu_n}_{(n-1)}
\end{eqnarray}
which allows us to recursively map the three $\gamma$ strings of
$G_{(\alpha^\prime \beta^\prime \gamma^\prime | \alpha \beta \gamma)}$ to the
$\Gamma_{(n)}$ basis. In particular the three $\gamma$ strings can be written
in terms of the objects 
\begin{equation}
\Gamma_{qrs ( \alpha\alpha^\prime | \beta\beta^\prime |
\gamma\gamma^\prime)} ~=~
\Gamma_{(q) \alpha\alpha^\prime} \otimes \Gamma_{(r) \beta\beta^\prime} \otimes
\Gamma_{(s) \gamma\gamma^\prime} 
\end{equation}
where $q$, $r$ and $s$ will be even from the fact that this is a massless 
computation. The spinor indices are included for reference when it comes to 
their multiplication for instance but will omitted hereafter. 

In indicating this spinor structure the Lorentz indices have been suppressed as
it is not possible to easily accommodate them in a general way. The explicit 
Lorentz contracted versions that actually arise to four loops will be provided 
later. Not only will there be contractions of indices across strings but the 
$\Gamma_{qrs}$ can contain at most two external momenta with each on a separate 
string due to the antisymmetry of the $\Gamma_{(n)}$ matrices. Such external 
momenta dependent $\Gamma_{qrs}$ matrices cannot be present in the final 
renormalization constant for ${\cal O}_{\alpha \beta \gamma}^{ijk}$ since the 
operator associated with them would invalidate renormalizability. They can be 
and are present in the evaluation of an individual graph but have to cancel 
upon summation of all the graphs. As an orientation point we have provided an 
example of a graph which contains $\Gamma_{484}$ in Figure \ref{figop} where
the labelling is from the left in the graph. The convention is that the 
subscripts count the number of $\gamma$ matrices in each quark line starting 
from the left. There will be other $\Gamma_{qrs}$ matrices in this graph due to
contractions but with lower subscript values.

{\begin{figure}[ht]
\begin{center}
\includegraphics[width=7.5cm,height=7.5cm]{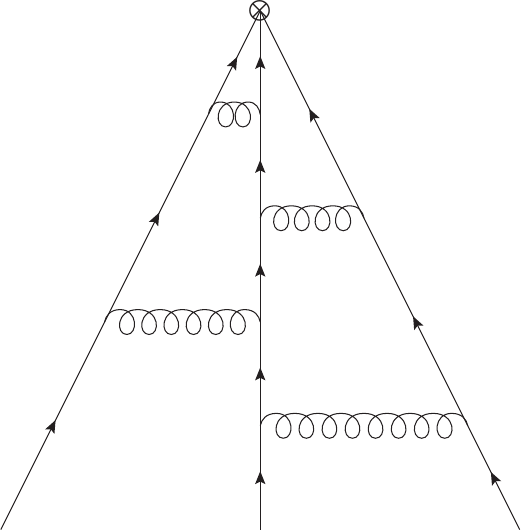}
\end{center}
\caption{Four loop graph contribution to
$G_{(\alpha^\prime \beta^\prime \gamma^\prime | 
\alpha \beta \gamma)}^{(i^\prime j^\prime k^\prime | ijk)}(p)$ containing the
$\gamma$ matrix structure $\Gamma_{484}$.}
\label{figop}
\end{figure}}

While the $\Gamma_{qrs}$ matrices form the core structure of the $\gamma$ 
strings of a graph a more general combination emerges in the renormalization 
constant of ${\cal O}^{ijk}$. These are denoted by $\Cc_{qrs}$ and are 
symmetric in the even integers $q$, $r$ and $s$. In particular there are three 
classes of $\Cc_{qrs}$ given explicitly by
\begin{eqnarray}
\Cc_{qqq} &\equiv& \Gamma_{qqq} ~~~,~~~
\Cc_{qqr} ~\equiv~ \Gamma_{qqr} ~+~ \Gamma_{qrq} ~+~ \Gamma_{rqq} ~=~ 
\Cc_{qrq} ~=~ \Cc_{rqq} \nonumber \\
\Cc_{qrs} &\equiv& \Gamma_{qrs} ~+~ \Gamma_{qsr} ~+~ \Gamma_{rqs} ~+~ 
\Gamma_{rsq} ~+~ \Gamma_{srq} ~+~ \Gamma_{sqr} 
\end{eqnarray}
where there is no summation over repeated labels, $q$, $r$ and $s$ are all 
different here and we will use the convention that the integers in the explicit
label of a particular $\Cc_{qrs}$ will be in descending order. To four loops 
the objects that are present in the renormalization constant are 
\begin{eqnarray}
\Cc_{000} &=& \Gamma_{000} ~~,~~ 
\Cc_{220} ~=~ \Gamma_{220} ~+~ \Gamma_{202} ~+~ \Gamma_{022} \nonumber \\
\Cc_{440} &=& \Gamma_{440} ~+~ \Gamma_{404} ~+~ \Gamma_{044} ~~,~~
\Cc_{422} ~=~ \Gamma_{422} ~+~ \Gamma_{242} ~+~ \Gamma_{224} \nonumber \\
\Cc_{660} &=& \Gamma_{660} ~+~ \Gamma_{606} ~+~ \Gamma_{066} ~~,~~ 
\Cc_{444} ~=~ \Gamma_{444} \nonumber \\
\Cc_{642} &=& \Gamma_{642} ~+~ \Gamma_{624} ~+~ \Gamma_{462} ~+~ \Gamma_{426} 
~+~ \Gamma_{246} ~+~ \Gamma_{264} \nonumber \\
\Cc_{880} &=& \Gamma_{880} ~+~ \Gamma_{808} ~+~ \Gamma_{088} \nonumber \\
\Cc_{862} &=& \Gamma_{862} ~+~ \Gamma_{826} ~+~ \Gamma_{682} ~+~ \Gamma_{628} 
~+~ \Gamma_{268} ~+~ \Gamma_{286} \nonumber \\
\Cc_{844} &=& \Gamma_{844} ~+~ \Gamma_{484} ~+~ \Gamma_{448} ~~,~~
\Cc_{664} ~=~ \Gamma_{664} ~+~ \Gamma_{646} ~+~ \Gamma_{466} ~. 
\end{eqnarray}
However at intermediate steps
\begin{eqnarray}
\Cc_{222} &=& \Gamma_{222} ~~,~~ 
\Cc_{442} ~=~ \Gamma_{442} ~+~ \Gamma_{424} ~+~ \Gamma_{244} \nonumber \\
\Cc_{662} &=& \Gamma_{662} ~+~ \Gamma_{626} ~+~ \Gamma_{266} ~~,~~ 
\Cc_{644} ~=~ \Gamma_{644} ~+~ \Gamma_{464} ~+~ \Gamma_{446} 
\end{eqnarray}
will also arise but when all the graphs are summed they are absent. We note
that for each of these the sum of the $q$, $r$ and $s$ labels is not a multiple
of four. Within each $\Cc_{qrs}$ definition the Lorentz index contractions are
\begin{eqnarray}
\Gamma_{000} &=& \Gamma_{(0)} \otimes \Gamma_{(0)} \otimes \Gamma_{(0)}
\nonumber \\
\Gamma_{220} &=& \Gamma_{(2)}^{\mu_1\mu_2} \otimes 
\Gamma_{(2) \, \mu_1\mu_2} \otimes \Gamma_{(0)}
\nonumber \\
\Gamma_{440} &=& \Gamma_{(4)}^{\mu_1\mu_2\mu_3\mu_4} \otimes 
\Gamma_{(4) \, \mu_1\mu_2\mu_3\mu_4} \otimes \Gamma_{(0)}
\nonumber \\
\Gamma_{422} &=& \Gamma_{(4)}^{\mu_1\mu_2\mu_3\mu_4} \otimes 
\Gamma_{(2) \, \mu_1\mu_2} \otimes \Gamma_{(2) \, \mu_3\mu_4} 
\nonumber \\
\Gamma_{660} &=& \Gamma_{(6)}^{\mu_1\mu_2\mu_3\mu_4\mu_5\mu_6} \otimes 
\Gamma_{(6) \, \mu_1\mu_2\mu_3\mu_4\mu_5\mu_6} \otimes \Gamma_{(0)}
\nonumber \\
\Gamma_{444} &=& \Gamma_{(4)}^{\mu_1\mu_2\mu_3\mu_4} \otimes 
\Gamma_{(4) \, \mu_1\mu_2}^{~~~~~~~~~\mu_5\mu_6} \otimes 
\Gamma_{(4) \, \mu_3\mu_4\mu_5\mu_6} 
\nonumber \\
\Gamma_{642} &=& \Gamma_{(6)}^{\mu_1\mu_2\mu_3\mu_4\mu_5\mu_6} \otimes 
\Gamma_{(4) \, \mu_1\mu_2\mu_3\mu_4} \otimes 
\Gamma_{(2) \, \mu_5\mu_6} \nonumber \\
\Gamma_{880} &=& \Gamma_{(8)}^{\mu_1\mu_2\mu_3\mu_4\mu_5\mu_6\mu_7\mu_8} 
\otimes 
\Gamma_{(8) \, \mu_1\mu_2\mu_3\mu_4\mu_5\mu_6\mu_7\mu_8} \otimes \Gamma_{(0)}
\nonumber \\
\Gamma_{862} &=& \Gamma_{(8)}^{\mu_1\mu_2\mu_3\mu_4\mu_5\mu_6\mu_7\mu_8} 
\otimes \Gamma_{(6) \, \mu_1\mu_2\mu_3\mu_4\mu_5\mu_6} \otimes 
\Gamma_{(2) \, \mu_7\mu_8} \nonumber \\
\Gamma_{844} &=& \Gamma_{(8)}^{\mu_1\mu_2\mu_3\mu_4\mu_5\mu_6\mu_7\mu_8} 
\otimes \Gamma_{(4) \, \mu_1\mu_2\mu_3\mu_4} \otimes 
\Gamma_{(4) \, \mu_5\mu_6\mu_7\mu_8} \nonumber \\
\Gamma_{664} &=& \Gamma_{(6)}^{\mu_1\mu_2\mu_3\mu_4\mu_5\mu_6} 
\otimes \Gamma_{(6) \, \mu_1\mu_2\mu_3\mu_4\mu_7\mu_8} \otimes 
\Gamma_{(4) \, \mu_5\mu_6}^{~~~~~~~~~\mu_7\mu_8} 
\end{eqnarray}
for those $\gamma$ strings that arise in the renormalization appearing for the
first time at $L$ loops where $q$~$+$~$r$~$+$~$s$~$=$~$4L$. For the remainder
\begin{eqnarray}
\Gamma_{222} &=& \Gamma_{(2)}^{\mu_1\mu_2} \otimes 
\Gamma_{(2) \, \mu_1\mu_3} \otimes \Gamma_{(2) \, \mu_2}^{~~~~~~\mu_3} 
\nonumber \\
\Gamma_{442} &=& \Gamma_{(4)}^{\mu_1\mu_2\mu_3\mu_4} \otimes 
\Gamma_{(4) \, \mu_1\mu_2\mu_3}^{~~~~~~~~~~~\mu_5} \otimes 
\Gamma_{(2) \, \mu_4\mu_5} 
\nonumber \\
\Gamma_{662} &=& \Gamma_{(6)}^{\mu_1\mu_2\mu_3\mu_4\mu_5\mu_6} 
\otimes \Gamma_{(6) \, \mu_1\mu_2\mu_3\mu_4\mu_5\mu_7} \otimes 
\Gamma_{(2) \, \mu_6}^{~~~~~~\mu_7} ~\nonumber \\
\Gamma_{644} &=& \Gamma_{(6)}^{\mu_1\mu_2\mu_3\mu_4\mu_5\mu_6} 
\otimes \Gamma_{(4) \, \mu_1\mu_2\mu_3\mu_7} \otimes 
\Gamma_{(4) \, \mu_4\mu_5\mu_6}^{~~~~~~~~~~~\mu_7} 
\end{eqnarray}
where the contractions for the permuted distinct $q$, $r$ and $r$ labels are 
deduced in an obvious fashion. In addition there are $\gamma$ strings which 
involve contractions with the external momenta $p$ such as
\begin{eqnarray}
\Gamma_{2p2p2} &=&
\Gamma_{(2)}^{p \mu_1} \otimes \Gamma_{(2) \, p \mu_2} \otimes 
\Gamma_{(2) \, \mu_1}^{~~~~~~ \mu_2}
\nonumber \\
\Gamma_{22p4p} &=&
\Gamma_{(2)}^{\mu_1 \mu_2} \otimes \Gamma_{(2)}^{p \mu_3} \otimes 
\Gamma_{(4) \, p \mu_1 \mu_2 \mu_3}
\nonumber \\
\Gamma_{24p4p} &=&
\Gamma_{(2)}^{\mu_1 \mu_2} \otimes \Gamma_{(4) \, p \mu_1 \mu_3 \mu_4} \otimes 
\Gamma_{(4) \, p \mu_2}^{~~~~~~~ \mu_3 \mu_4}
\nonumber \\
\Gamma_{2p4p4} &=&
\Gamma_{(2)}^{p \mu_1} \otimes \Gamma_{(4)}^{p \mu_2 \mu_3 \mu_4} \otimes 
\Gamma_{(4) \, \mu_1 \mu_2 \mu_3 \mu_4}
\nonumber \\
\Gamma_{4p4p4} &=&
\Gamma_{(4) \, p \mu_1 \mu_2 \mu_5} \otimes 
\Gamma_{(4) \, p \mu_3 \mu_4}^{~~~~~~~~~~ \mu_5} \otimes 
\Gamma_{(4)}^{\mu_1 \mu_2 \mu_3 \mu_4}
\nonumber \\
\Gamma_{24p6p} &=&
\Gamma_{(2)}^{\mu_1 \mu_2} \otimes \Gamma_{(4)}^{p \mu_3 \mu_4 \mu_5} \otimes 
\Gamma_{(6) \, p \mu_1 \mu_2 \mu_3 \mu_4 \mu_5}
\nonumber \\
\Gamma_{26p6p} &=&
\Gamma_{(2)}^{\mu_1 \mu_2} \otimes 
\Gamma_{(6) \, p \mu_1 \mu_3 \mu_4 \mu_5 \mu_6} \otimes 
\Gamma_{(6) \, p \mu_2}^{~~~~~~~ \mu_3 \mu_4 \mu_5 \mu_6}
\nonumber \\
\Gamma_{4p4p6} &=&
\Gamma_{(4) \, p \mu_1 \mu_2 \mu_3} \otimes \Gamma_{(4) \, p \mu_4 \mu_5 \mu_6}
\otimes \Gamma_{(6)}^{\mu_1 \mu_2 \mu_3 \mu_4 \mu_5 \mu_6}
\nonumber \\
\Gamma_{46p6p} &=&
\Gamma_{(4)}^{\mu_1 \mu_2 \mu_3 \mu_4} \otimes 
\Gamma_{(6) \, p \mu_1 \mu_2 \mu_5 \mu_6 \mu_7} \otimes 
\Gamma_{(6) \, p \mu_3 \mu_4}^{~~~~~~~~~~ \mu_5 \mu_6 \mu_7}
\nonumber \\
\Gamma_{npnp0} &=&
\Gamma_{(n)}^{p \mu_1 \ldots \mu_{n-1}} \otimes 
\Gamma_{(n) \, p \mu_1 \ldots \mu_{n-1}} \otimes \Gamma_{(0)} 
\end{eqnarray}
where $n$ is even and we use the convention that the appearance of $p$ in the 
index string corresponds to its contraction with a Lorentz index. In the 
overall determination of the operator renormalization constants the poles in 
$\epsilon$ associated with these structures cancel. Otherwise the original
operator would be non-renormalizable as they would have to have a factor of 
$1/p^2$ on dimensional grounds and therefore would correspond to a non-local 
$3$-quark operator.

Having introduced the compact notation for the underlying $\gamma$ matrix
structure that is associated with (\ref{gfdef}) we can now introduce the 
renormalization constant for ${\cal O}^{ijk}$. Denoting bare quantities with
the subscript ${}_{\bare}$ then 
\begin{equation}
{\cal O}^{ijk}_{\bare \, \alpha\beta\gamma} ~=~ 
Z_{\alpha ~ \beta ~~ \gamma}^{~\alpha^\prime ~ \beta^\prime \, \gamma^\prime}
{\cal O}^{ijk}_{\alpha^\prime\beta^\prime\gamma^\prime} ~.
\end{equation}
Explicit computations indicate the renormalization constant takes the form
\begin{equation}
Z_{\alpha\alpha^\prime\beta\beta^\prime\gamma\gamma^\prime} ~=~
\Cc_{000 (\alpha\alpha^\prime | \beta\beta^\prime | \gamma\gamma^\prime)} ~+~ 
\sum_{{\cal K}} a_{qrs}(\epsilon) \, 
\Cc_{qrs (\alpha\alpha^\prime | \beta\beta^\prime | \gamma\gamma^\prime)} 
\end{equation}
where the indexing set ${\cal K}$ ranges over all valid combinations of $q$, 
$r$ and $s$ of the basic $\Gamma_{(n)}$ matrix structure and thus there is no 
sum over individual $q$, $r$ and $s$. The $\epsilon$ dependent poles are 
contained in
\begin{equation}
a_{qrs}(\epsilon) ~=~ \sum_{n=1}^\infty \frac{a^{(n)}_{qrs}}{\epsilon^n} 
\end{equation}
where we have chosen to renormalize in the $\MSbar$ scheme. To extract the
particular residues we have evaluated all the Feynman graphs to four loops
contributing to (\ref{gfdef}) for the momentum configuration of Figure 
\ref{figgf}. As this is in effect a $2$-point function the {\sc Forcer} 
algorithm of \cite{15,16} can be applied. The Feynman graphs are generated with
{\sc Qgraf} \cite{33} and there are $3$, $40$, $784$ and $19061$ graphs from 
one to four loops respectively. A central aspect of the evaluation is the 
mapping of the $\gamma$ matrix structures that arise to the $\Cc_{qrs}$ basis 
in a way that is algebraically manageable for which {\sc Form} was the ideal 
tool.

\sect{Results.}

Having described the basic structures of the renormalization process for
(\ref{gfdef}) we have extracted the four loop renormalization constant for 
${\cal O}^{ijk}_{\alpha\beta\gamma}$ in terms of the general $\Cc_{qrs}$
basis in the $\MSbar$ scheme. Its conversion to the associated operator
anomalous dimension is effected via the spinor equation
\begin{equation}
\gamma_{{\cal O} \, \alpha\alpha^{\prime\prime} \beta\beta^{\prime\prime}
\gamma\gamma^{\prime\prime}}(a) 
Z_{{\cal O} \, ~~ \alpha^\prime ~~ \beta^\prime ~~ 
\gamma^\prime}^{~~\alpha^{\prime \prime} ~\, \beta^{\prime \prime} ~\, 
\gamma^{\prime \prime}} ~=~
Z_{{\cal O} \, \alpha\alpha^{\prime\prime} \beta\beta^{\prime\prime}
\gamma\gamma^{\prime\prime}} 
\gamma_{{\cal O} \, ~~ \alpha^\prime ~~ \beta^\prime ~~ 
\gamma^\prime}^{~~\alpha^{\prime \prime} ~\, \beta^{\prime \prime} ~\, 
\gamma^{\prime \prime}}(a) ~=~ \beta(a) \frac{\partial }{\partial a}
Z_{{\cal O} \, \alpha\alpha^{\prime} \beta\beta^{\prime}
\gamma\gamma^{\prime}} 
\label{ztogamma}
\end{equation}
which is the functional generalization of the usual derivative of the logarithm
of a renormalization constant and where $a$ is the coupling constant. The 
operator anomalous dimension has to be defined in this manner since the 
logarithm of a matrix only has meaning in a perturbative or Taylor series 
sense. The equivalence of (\ref{ztogamma}) to the canonical definition of an 
anomalous dimension is preserved by the separate left and right 
multiplications. Both have to lead to the same value of $\gamma_{\cal O}(a)$. 
The final stage of extracting the anomalous dimension involves the 
multiplications on the left side of (\ref{ztogamma}) which involve products of 
$\Cc_{qrs}$. We have provided the relations that were needed to four loops in 
Appendix A. As the labels of $\Cc_{qrs}$ that contribute to the anomalous
dimension sum to multiples of $4n$ for positive integers $n$ the sum of the 
labels of the product of two $\Cc_{qrs}$ needed for a four loop renormalization
must be no more than $16$ which is $4$ times the loop order. Therefore there 
are $1$, $2$ and $6$ relevant products at two, three and four loops 
respectively excluding those involving the identity $\Cc_{000}$. Consequently 
we find the anomalous dimension is
\begin{eqnarray}
\gamma_{\cal O}^{\mbox{\scriptsize{naive}}}(a) &=&
\left[ 
\frac{1}{3} a
+ \left[
\frac{49}{18}
- \frac{1}{27} \Nf
\right] a^2
+ \left[
\frac{5479}{108}
- \frac{212}{81} \Nf
- \frac{127}{9} \zeta_3
- \frac{40}{9} \zeta_3 \Nf
- \frac{13}{81} \Nf^2
\right] a^3
\right. \nonumber \\
&& \left. ~
+ \left[
\frac{16}{81} \zeta_3 \Nf^3
+ \frac{440}{81} \zeta_3 \Nf^2
+ \frac{533}{9} \zeta_4 \Nf
+ \frac{1397}{6} \zeta_4
+ \frac{1705}{1458} \Nf^2
+ \frac{5380}{27} \zeta_5 \Nf
\right. \right. \nonumber \\
&& \left. \left. ~~~~~
+ \frac{2320925}{1944}
- \frac{300335}{486} \zeta_5
- \frac{266069}{972} \zeta_3
- \frac{150629}{972} \Nf
- \frac{39731}{162} \zeta_3 \Nf
- \frac{40}{9} \zeta_4 \Nf^2
\right. \right. \nonumber \\
&& \left. \left. ~~~~~
- \frac{5}{27} \Nf^3
\right] a^4
\right] \Cc_{220}
\nonumber \\
&& 
+ \left[ 
\left[
4 \Nf
- 70
\right] a^2
+ \left[
 \frac{434}{3} \zeta_3
+ \frac{1708}{9} \Nf
- \frac{16492}{9}
- \frac{20}{9} \Nf^2
\right] a^3
\right. \nonumber \\
&& \left. ~~~~
+ \left[
\frac{434}{3} \zeta_4 \Nf
+ \frac{5555}{27} \zeta_5
+ \frac{14900}{27} \zeta_3 \Nf
+ \frac{172582}{27} \Nf
+ \frac{212572}{27} \zeta_3
- 2387 \zeta_4
\right. \right. \nonumber \\
&& \left. \left. ~~~~~~~~
- 280 \zeta_5 \Nf
- \frac{374575}{9}
- \frac{1285}{9} \Nf^2
- \frac{160}{3} \zeta_3 \Nf^2
- \frac{140}{81} \Nf^3
\right] a^4
\right] \Cc_{000}
\nonumber \\
&& 
+ \left[ 
- \frac{1}{9} a^2
+ \left[
\frac{2}{81} \Nf
+ \frac{13}{36} \zeta_3
+ \frac{127}{216}
\right] a^3
\right. \nonumber \\
&& \left. ~~~~
+ \left[
\frac{13}{36} \zeta_4 \Nf
+ \frac{299}{2916} \Nf^2
+ \frac{829}{324} \zeta_3 \Nf
+ \frac{3287}{1944} \Nf
+ \frac{97705}{1944} \zeta_5
- \frac{45415}{1944} \zeta_3
\right. \right. \nonumber \\
&& \left. \left. ~~~~~~~~
- \frac{13045}{324}
- \frac{143}{24} \zeta_4
- \frac{5}{3} \zeta_5 \Nf
\right] a^4
\right] \Cc_{440}
\nonumber \\
&& 
+ \left[ 
\frac{1}{18} a^2
+ \left[
\frac{1}{162} \Nf
+ \frac{19}{27}
+ \frac{25}{18} \zeta_3
\right] a^3
\right. \nonumber \\
&& \left. ~~~~
+ \left[
\frac{25}{18} \zeta_4 \Nf
+ \frac{695}{72}
+ \frac{5549}{243} \zeta_3
- \frac{15545}{972} \zeta_5
- \frac{1127}{972} \Nf
- \frac{275}{12} \zeta_4
- \frac{77}{81} \zeta_3 \Nf
\right. \right. \nonumber \\
&& \left. \left. ~~~~~~~~
- \frac{40}{9} \zeta_5 \Nf
- \frac{37}{729} \Nf^2
\right] a^4
\right] \Cc_{422}
\nonumber \\
&& 
+ \left[ 
\left[
\frac{59}{432} \zeta_3
- \frac{101}{1296}
\right] a^3
\right. \nonumber \\
&& \left. ~~~~
+ \left[
\frac{59}{432} \zeta_4 \Nf
+ \frac{10535}{7776}
+ \frac{21293}{1944} \zeta_3
- \frac{128585}{7776} \zeta_5
- \frac{649}{288} \zeta_4
- \frac{389}{11664} \Nf
\right. \right. \nonumber \\
&& \left. \left. ~~~~~~~~
- \frac{295}{3888} \zeta_3 \Nf
\right] a^4
\right] \Cc_{660}
\nonumber \\
&& 
+ \left[
- \frac{17}{216} a^3
+ \left[
\frac{23}{162} \zeta_3 \Nf
+ \frac{1087}{1296} \zeta_3
- \frac{6775}{1296} \zeta_5
- \frac{6727}{1296}
- \frac{257}{1944} \Nf
\right] a^4
\right] \Cc_{444}
\nonumber \\
&& 
+ \left[
\left[
\frac{1}{16} \zeta_3
- \frac{1}{16}
\right] a^3
\right. \nonumber \\
&& \left. ~~~~
+ \left[
\frac{1}{16} \zeta_4 \Nf
+ \frac{9037}{7776}
+ \frac{9295}{648} \zeta_3
- \frac{47675}{2592} \zeta_5
- \frac{125}{3888} \Nf
- \frac{33}{32} \zeta_4
- \frac{5}{144} \zeta_3 \Nf
\right] a^4
\right] \Cc_{642}
\nonumber \\
&& 
+ \left[
\left[
\frac{25}{972} \zeta_5
+ \frac{31}{15552}
- \frac{1411}{15552} \zeta_3
\right] \Cc_{880}
+ \left[
\frac{205}{1944} \zeta_5
- \frac{247}{7776} \zeta_3
- \frac{113}{1944}
\right] \Cc_{862}
\right. \nonumber \\
&& \left. ~~~~
+ \left[
\frac{149}{2592} \zeta_3
- \frac{35}{3888}
- \frac{5}{48} \zeta_5
\right] \Cc_{844}
+ \left[
\frac{55}{216} \zeta_5
- \frac{61}{864} \zeta_3
- \frac{29}{216}
\right] \Cc_{664}
\right] a^4 \nonumber \\
&& +~ O(a^5) ~.
\label{anomnaive}
\end{eqnarray}
where $\zeta_n$ is the Riemann zeta function. In terms of conventions for this 
anomalous dimension we follow those of \cite{20} where the one loop term has a 
factor of $\third$ in contrast to the convention used in \cite{13}. It is 
evident that no $\Cc_{qrs}$ appears before loop order $L$ where 
$L$~$=$~$\quarter(q+r+s)$ here which acts as an accounting check. Numerically 
(\ref{anomnaive}) is
\begin{eqnarray}
\gamma_{\cal O}^{\mbox{\scriptsize{naive}}}(a) &=&
0.333333 \Cc_{220} a 
\nonumber \\
&&
+ \left[
\left[ 4.000000 \Nf - 70.000000 \right] \Cc_{000}
+ \left[
2.722222
- 0.037037 \Nf 
\right] \Cc_{220} 
+ 0.055556 \Cc_{422}
\right. \nonumber \\
&& \left. ~~~~
-~ 0.111111 \Cc_{440}
\right] a^2 
\nonumber \\
&&
+ \left[ 
\left[ 
189.777778 \Nf 
- 2.222222 \Nf^2 
- 1658.546879
\right] \Cc_{000} 
\right. \nonumber \\
&& \left. ~~~~
+ \left[
33.769123
- 0.160494 \Nf^2 
- 7.959759 \Nf 
\right] \Cc_{220} 
\right. \nonumber \\
&& \left. ~~~~
+ \left[
0.006173 \Nf 
+ 2.373227
\right] \Cc_{422} 
+ \left[
0.024691 \Nf 
+ 1.022039
\right] \Cc_{440} 
+ 0.012629 \Cc_{642}
\right. \nonumber \\
&& \left. ~~~~
+~ 0.086238 \Cc_{660}
- 0.078704 \Cc_{444}
\right] a^3 
\nonumber \\
&&
+ \left[ 
\left[
6921.519577 \Nf 
- 1.728395 \Nf^3 
- 206.887479 \Nf^2 
- 34525.773200 \right] \Cc_{000}
\right. \nonumber \\
&& \left. ~~~~
+ \left[
0.052258 \Nf^3 
+ 2.888776 \Nf^2 
- 179.061229 \Nf 
+ 476.055481 
\right] \Cc_{220}
\right. \nonumber \\
&& \left. ~~~~
+ \left[
- 0.050754 \Nf^2 
- 5.407502 \Nf 
- 4.284400
\right] \Cc_{422}
+ 0.044906 \Cc_{664}
\right. \nonumber \\
&& \left. ~~~~
+ \left[
0.102538 \Nf^2 
+ 3.429103 \Nf 
- 22.677437
\right] \Cc_{440}
+ \left[
0.038461 \Nf 
- 9.603048
\right] \Cc_{444}
\right. \nonumber \\
&& \left. ~~~~
+ \left[
- 0.006243 \Nf 
- 1.783863
\right] \Cc_{642}
+ \left[
0.023261 \Nf 
- 5.064598
\right] \Cc_{660}
\right. \nonumber \\
&& \left. ~~~~
-~ 0.047916 \Cc_{844}
+ 0.013037 \Cc_{862}
- 0.080397 \Cc_{880}
\right] a^4 ~+~ O(a^5) ~.
\end{eqnarray}

We have given the label naive to the anomalous dimension of (\ref{anomnaive})
for a particular reason. It is the anomalous dimension of ${\cal O}$ in the
regularized theory with $\epsilon$~$\neq$~$0$. Ordinarily the passage back to
four dimensions is straightforward in the limit of $\epsilon$~$\to$~$0$.
However this limit cannot be taken immediately since (\ref{anomnaive}) contains
evanescent $\Cc_{qrs}$ structures which correspond to any one of of $q$, $r$ or
$s$ being strictly more than $4$ as is apparent from (\ref{gamndto4}). To 
extract an anomalous dimension for ${\cal O}$ in four dimensions requires an 
additional procedure which was introduced in \cite{13} and motivated by a
relation used in the two loop computation of \cite{12}. The basic idea at two 
loops was to realize that not all the $\Cc_{qrs}$ which appear are independent 
since, for instance, 
\begin{equation}
\Cc_{422} ~=~ -~ 3 d(d-1) \Cc_{000} ~-~ 2 (d-3) \Cc_{220} ~-~ 
\frac{1}{2} \Cc_{440} ~+~ \frac{1}{2} \Cc_{220}^2 
\label{gaml2reln}
\end{equation} 
was derived in \cite{12}. While $\Cc_{422}$ is not related to an evanescent 
structure it was expressed in terms of the other $\Cc_{qrs}$ since the
eigenvalues of $\Cc_{220}$ and $\Cc_{440}$ could be determined for four 
dimensional baryon operators \cite{12} which we will recall later. However
similar $d$-dimensional relations can be established for evanescent $\Cc_{qrs}$
structures and three loop ones were given in \cite{13} which were
\begin{eqnarray}
\Cc_{660} &=& -~ 12 d(d-1)(2d-1) \Cc_{000} ~-~ 3 (d-1)(7d-24) \Cc_{220} ~-~ 6 (2d-5) \Cc_{440} 
\nonumber \\
&& +~ 2 (3d-4) \Cc_{220}^2 ~-~ \frac{1}{2} \Cc_{220}^3 ~+~ \frac{3}{2} \Cc_{220} \Cc_{440} ~+~ 
3 \Cc_{444} \nonumber \\
\Cc_{642} &=& 12 d(d-1)(2d-7) \Cc_{000} ~+~ 9 (d^2-9d+16) \Cc_{220} ~+~ 2 (2d-5) \Cc_{440} 
\nonumber \\ 
&& -~ 2 (3d-10) \Cc_{220}^2 ~+~ \frac{1}{2} \Cc_{220}^3 ~-~ 
\frac{1}{2} \Cc_{220} \Cc_{440} ~-~ 3 \Cc_{444} ~. 
\label{gaml3reln}
\end{eqnarray} 
The procedure to establish these relations is clear. The two loop relation of 
(\ref{gaml2reln}) involves the square of the one loop structure which is
$\Cc_{220}$. In fact evaluating the square was the starting point to deduce 
(\ref{gaml2reln}). At three loops to produce the evanescent structures there 
are two possible combinations of $\Cc_{220}$ and $\Cc_{440}$ that will lead to 
the two three loop evanescent $\Cc_{qrs}$ structures. These are $\Cc_{220}^3$ 
and $\Cc_{220} \Cc_{440}$ which both have a total $(q+r+s)$ value of $12$. 
Evaluating these lead to the relations of (\ref{gaml3reln}). 

Therefore the algorithm is set to construct the relations for the four loop 
evanescent $\Cc_{qrs}$ structures. Since there are four such objects then we
require four combinations of lower order non-evanescent $\Cc_{qrs}$ objects.
Clearly the starting point for these are $\Cc_{220}^4$, $\Cc_{220}^2 
\Cc_{440}$, $\Cc_{440}^2$ and $\Cc_{220} \Cc_{444}$ as $\Cc_{444}$ is 
non-evanescent and appears first at three loops. Evaluating these products 
produces four linear combinations of $\Cc_{880}$, $\Cc_{862}$, $\Cc_{844}$ and 
$\Cc_{664}$ and inverting them gives 
\begin{eqnarray}
\Cc_{880} &=&
6 d (d - 1) (69 d^2 - 437 d + 280) \Cc_{000}
- \frac{1}{2} \Cc_{220}^4 
+ 16 (d - 4) \Cc_{220}^3 
+ \Cc_{220}^2 \Cc_{440}
\nonumber \\
&&
-~ 2 (65 d^2 - 437 d + 512) \Cc_{220}^2 
- 8 (4 d - 19) \Cc_{220} \Cc_{440}
+ 4 \Cc_{220} \Cc_{444}
- 24 (3 d - 22) \Cc_{444}
\nonumber \\
&&
+~ 16 (16 d^3 - 183 d^2 + 545 d - 384) \Cc_{220}
+ \frac{1}{2} \Cc_{440}^2 
+ 6 (23 d^2 - 175 d + 256) \Cc_{440}
\nonumber \\
\Cc_{862} &=&
-~ 6 d (d - 1) (21 d^2 - 245 d + 556) \Cc_{000}
- (4 d - 35) \Cc_{220}^2 
+ \frac{1}{2} \Cc_{220}^2 \Cc_{440}
+ 6 (6 d - 67) \Cc_{444}
\nonumber \\
&&
+~ ( 37 d^2 - 433 d + 976 ) \Cc_{220}^2
+ ( 2 d - 29 ) \Cc_{220} \Cc_{440}
- \Cc_{220} \Cc_{444}
\nonumber \\
&&
-~ 2 (14 d^3 - 303 d^2 + 1801 d - 2640) \Cc_{220}
- \frac{1}{2} \Cc_{440}^2 
- 12 (2 d^2 - 18 d + 23) \Cc_{440}
\nonumber \\
\Cc_{844} &=&
-~ 3 d (d - 1) (17 d^2 - 81 d + 160) \Cc_{000}
+ \frac{1}{4} \Cc_{220}^4 
- 4 (d - 2) \Cc_{220}^3
- \frac{1}{2} \Cc_{220}^2 \Cc_{440}
\nonumber \\
&&
+~ ( 17 d^2 - 85 d + 128 ) \Cc_{220}^2 
+ 4 (d - 1) \Cc_{220} \Cc_{440}
- 2 \Cc_{220} \Cc_{444}
+ 12 (d + 8) \Cc_{444}
\nonumber \\
&&
-~ 8 (d^3 - 18 d^2 + 71 d - 96) \Cc_{220}
+ \frac{1}{4} \Cc_{440}^2 
- 3 (3 d^2 - 11 d + 24) \Cc_{440}
\nonumber \\
\Cc_{664} &=&
6 d (d - 1) (d^2 - d - 8) \Cc_{000}
+ \Cc_{220}^3 
- (d^2 + 3 d - 20) \Cc_{220}^2 
- \Cc_{220} \Cc_{440}
+ \Cc_{220} \Cc_{444}
\nonumber \\
&&
-~ 6 (2 d - 7) \Cc_{444}
+ 2 (2 d^3 - 15 d^2 + 13 d + 24) \Cc_{220}
+ d (d - 1) \Cc_{440}
\label{gaml4reln}
\end{eqnarray}
in $d$-dimensions. The right hand side of each of these and the lower order 
ones contain only $\Cc_{qrs}$ structures with individual $q$, $r$ or $s$ values
of $0$, $2$ or $4$. It is these relations that will allow us to construct an 
anomalous dimension for ${\cal O}$ in four dimensions. 

However it is worth recording the argument for the five and six loop cases in 
order to appreciate the subtlety of the above derivation. At five loops there 
are five evanescent objects and precisely five independent combinations of 
$\Cc_{220}$, $\Cc_{440}$ and $\Cc_{444}$ with $10$ contracted Lorentz indices. 
The five relations can be constructed from the basis mapping
\begin{equation}
\{ \Cc_{10\,10\,0}, \Cc_{10\,82}, \Cc_{10\,64}, \Cc_{884}, 
\Cc_{866} \} ~\leftrightarrow~
\{ \Cc_{220}^5, \Cc_{220}^3 \Cc_{440}, \Cc_{220} \Cc_{440}^2,
\Cc_{220}^2 \Cc_{444}, \Cc_{440} \Cc_{444} \} ~.
\end{equation}
At six loops the number of evanescent structures and independent combinations
is seven. Specifically
\begin{eqnarray}
&& \{ \Cc_{12\,12\,0}, \Cc_{12\,10\,2}, \Cc_{12\,84}, \Cc_{12\,66}, 
\Cc_{10\,10\,4}, \Cc_{10\,86}, \Cc_{888} \} \nonumber \\
&& ~\leftrightarrow~
\{ \Cc_{220}^6, \Cc_{220}^4 \Cc_{440}, \Cc_{220}^2 \Cc_{440}^2, \Cc_{440}^3,
\Cc_{220}^3 \Cc_{444}, \Cc_{220} \Cc_{440} \Cc_{444}, \Cc_{444}^2 \} 
\end{eqnarray}
illustrates both sets. As the explicit relations similar to those of 
(\ref{gaml4reln}) are not required for the present computation we have not 
determined them here.

Having established the mapping of the evanescent $\Cc_{qrs}$ structures to the
four dimensional ones these are applied to the evaluation of the Green's
function {\em before} the renormalization constant for 
${\cal O}^{ijk}_{\alpha\beta\gamma}$ is extracted in the $\MSbar$ scheme
\cite{12,13}. In other words the $\epsilon$ expansion of the $d$-dependent 
coefficients of each term in (\ref{gaml4reln}) is effected first. This results 
in the four dimensional anomalous dimension
\begin{eqnarray}
\gamma_{\cal O}(a) &=&
\frac{1}{3} \Cc_{220} a
+ \left[
\left[
4 \Nf
- 72
\right] \Cc_{000}
+ \left[
\frac{47}{18}
- \frac{1}{27} \Nf
\right] \Cc_{220}
+ \frac{1}{36} \Cc_{220}^2 
- \frac{5}{36} \Cc_{440}
\right] a^2 
\nonumber \\
&&
+ \left[
\left[
\frac{1706}{9} \Nf
- 34 \zeta_3
- \frac{16094}{9}
- \frac{20}{9} \Nf^2
\right] \Cc_{000}
\right. \nonumber \\
&& \left. ~~~~
+ \left[
\frac{5873}{108}
- \frac{433}{18} \zeta_3
- \frac{71}{27} \Nf
- \frac{40}{9} \zeta_3 \Nf
- \frac{13}{81} \Nf^2
\right] \Cc_{220}
+ \left[
\frac{5}{648}
- \frac{1}{27} \zeta_3
\right] \Cc_{220}^3
\right. \nonumber \\
&& \left. ~~~~
+ \left[
\frac{1}{324} \Nf
- \frac{209}{324}
+ \frac{71}{27} \zeta_3
\right] \Cc_{220}^2
+ \left[
\frac{25}{144} \zeta_3
- \frac{37}{432}
\right] \Cc_{220} \Cc_{440}
\right. \nonumber \\
&& \left. ~~~~
+ \left[
\frac{91}{72}
- \frac{29}{12} \zeta_3
+ \frac{7}{324} \Nf
\right] \Cc_{440}
+ \left[
\frac{2}{9} \zeta_3
- \frac{1}{8}
\right] \Cc_{444}
\right] a^3 
\nonumber \\
&&
+ \left[
\left[
\frac{523475}{81} \Nf
- 120 \zeta_5 \Nf
- 34 \zeta_4 \Nf
+ 561 \zeta_4
- \frac{392305}{9}
- \frac{11417}{81} \Nf^2
- \frac{160}{3} \zeta_3 \Nf^2
\right. \right. \nonumber \\
&& \left. \left. ~~~~~
- \frac{140}{81} \Nf^3
+ \frac{227}{27} \zeta_3
+ \frac{5918}{9} \zeta_3 \Nf
+ \frac{412390}{27} \zeta_5
\right] \Cc_{000}
\right. \nonumber \\
&& \left. ~~~~
+ \left[
\frac{2252615}{1944}
- \frac{328151}{324} \zeta_3
- \frac{146083}{972} \Nf
- \frac{6463}{27} \zeta_3 \Nf
- \frac{40}{9} \zeta_4 \Nf^2
- \frac{5}{27} \Nf^3
+ \frac{16}{81} \zeta_3 \Nf^3
\right. \right. \nonumber \\
&& \left. \left. ~~~~~~~~
+ \frac{440}{81} \zeta_3 \Nf^2
+ \frac{887}{18} \zeta_4 \Nf
+ \frac{1853}{1458} \Nf^2
+ \frac{4763}{12} \zeta_4
+ \frac{5620}{27} \zeta_5 \Nf
+ \frac{145675}{324} \zeta_5
\right] \Cc_{220}
\right. \nonumber \\
&& \left. ~~~~
+ \left[
\frac{42383}{1296}
+ \frac{200135}{1944} \zeta_3
- \frac{208495}{972} \zeta_5
- \frac{5743}{5832} \Nf
- \frac{1507}{972} \zeta_3 \Nf
- \frac{781}{18} \zeta_4
- \frac{37}{1458} \Nf^2
\right. \right. \nonumber \\
&& \left. \left. ~~~~~~~~
- \frac{20}{9} \zeta_5 \Nf
+ \frac{71}{27} \zeta_4 \Nf
\right] \Cc_{220}^2
\right. \nonumber \\
&& \left. ~~~~
+ \left[
\frac{7}{11664} \Nf
- \frac{9821}{7776}
- \frac{1}{27} \zeta_4 \Nf
+ \frac{5}{243} \zeta_3 \Nf
+ \frac{11}{18} \zeta_4
+ \frac{2183}{3888} \zeta_3
+ \frac{4205}{1944} \zeta_5
\right] \Cc_{220}^3
\right. \nonumber \\
&& \left. ~~~~
+ \left[
\frac{929}{15552} \zeta_3
- \frac{605}{15552} \zeta_5
- \frac{101}{31104}
\right] \Cc_{220}^4
+ \left[
\frac{1015}{7776} \zeta_5
- \frac{2105}{15552} \zeta_3
- \frac{13}{576}
\right] \Cc_{220}^2 \Cc_{440}
\right. \nonumber \\
&& \left. ~~~~
+ \left[
\frac{2669}{972}
+ \frac{5513}{648} \zeta_3
- \frac{16165}{864} \zeta_5
- \frac{275}{96} \zeta_4
- \frac{125}{1296} \zeta_3 \Nf
- \frac{11}{324} \Nf
+ \frac{25}{144} \zeta_4 \Nf
\right] \Cc_{220} \Cc_{440}
\right. \nonumber \\
&& \left. ~~~~
+ \left[
\frac{173105}{648} \zeta_5
- \frac{98657}{1296}
- \frac{75593}{648} \zeta_3
- \frac{29}{12} \zeta_4 \Nf
+ \frac{5}{9} \zeta_5 \Nf
+ \frac{319}{8} \zeta_4
+ \frac{373}{2916} \Nf^2
\right. \right. \nonumber \\
&& \left. \left. ~~~~~~~~
+ \frac{679}{162} \zeta_3 \Nf
+ \frac{2603}{972} \Nf
\right] \Cc_{440}
+ \left[
\frac{895}{1944} \zeta_5
- \frac{2009}{3888} \zeta_3
- \frac{65}{1296}
\right] \Cc_{444} \Cc_{220} 
\right. \nonumber \\
&& \left. ~~~~
+ \left[
\frac{6721}{648}
- \frac{48235}{1296} \zeta_5
- \frac{18419}{1296} \zeta_3
- \frac{11}{3} \zeta_4
- \frac{11}{81} \Nf
+ \frac{1}{54} \zeta_3 \Nf
+ \frac{2}{9} \zeta_4 \Nf
\right] \Cc_{444}
\right. \nonumber \\
&& \left. ~~~~
+ \left[
\frac{865}{31104}
- \frac{1025}{15552} \zeta_5
- \frac{235}{15552} \zeta_3
\right] \Cc_{440}^2
\right] a^4 ~+~ O(a^5)
\label{gambargen}
\end{eqnarray}
to four loops or
\begin{eqnarray}
\gamma_{\cal O}(a) &=& 
\left[
0.333333 a 
+ ( 2.611111 - 0.037037 \Nf) a^2 
+ ( 25.463483 - 0.160494 \Nf^2 - 7.972105 \Nf ) a^3 
\right. \nonumber \\
&& \left. ~
+ (0.052258 \Nf^3 + 2.990285 \Nf^2 - 168.858883 \Nf + 837.104813) a^4 
\right] \Cc_{220}
\nonumber \\
&&
+ \left[
(4.000000 \Nf - 72.000000) a^2 
+ ( - 2.222222 \Nf^2 + 189.555556 \Nf - 1829.092157) a^3 
\right. \nonumber \\
&& \left. ~~~~
+ ( - 1.728395 \Nf^3 - 205.060319 \Nf^2 + 7091.843195 \Nf - 27134.427634) a^4
\right] \Cc_{000}
\nonumber \\
&&
+ \left[
- 0.138889 a^2 
+ (0.021605 \Nf - 1.6410819) a^3 
\right. \nonumber \\
&& \left. ~~~~
+ (0.127915 \Nf^2 + 5.676691 \Nf + 103.808552) a^4 
\right] \Cc_{440}
\nonumber \\
&&
+ \left[
0.027778 a^2 
+ (0.003086 \Nf + 2.515903) a^3 
\right. \nonumber \\
&& \left. ~~~~
+ ( - 0.025377 \Nf^2 - 2.306597 \Nf - 112.9280589) a^4 
\right] \Cc_{220}^2
\nonumber \\
&&
+ \left[
0.142124 a^3 
+ (0.126974 \Nf - 49.273223) a^4 
\right] \Cc_{444}
\nonumber \\
&&
+ \left[
0.123042 a^3 
+ (0.038014 \Nf - 9.528152) a^4 
\right] \Cc_{220} \Cc_{440}
\nonumber \\
&&
+ \left[
- 0.036805 a^3 
+ ( - 0.014752 \Nf + 2.316294) a^4 
\right] \Cc_{220}^3
\nonumber \\
&&
+ \left[ 
0.028219 \Cc_{220}^4 
- 0.058696 \Cc_{440}^2 
- 0.193887 \Cc_{444} \Cc_{220} 
- 0.049921 \Cc_{220}^2 \Cc_{440} 
\right] a^4 \nonumber \\
&& +~ O(a^5)
\end{eqnarray}
numerically.  One observation deserves comment and it concerns the comparison 
of the results of (\ref{anomnaive}) with (\ref{gambargen}). If one formally 
replaces the $\Cc_{qrs}$ matrices in (\ref{anomnaive}) with their strictly four
dimensional counterparts then (\ref{gambargen}) is reproduced. By this we mean 
the evanescent matrices of (\ref{anomnaive}) are replaced using the relations 
of (\ref{gaml3reln}) and (\ref{gaml4reln}) but with $d$ replaced by $4$. This 
is consistent with the general formal relation
\begin{equation}
\gamma_{\cal O}(a) ~=~ \lim_{d\to 4} 
\gamma_{\cal O}^{\mbox{\scriptsize{naive}}}(a) 
\label{gammalimit}
\end{equation}
which additionally provides a check on the translation of the two actual 
renormalization constants into their respective $\MSbar$ anomalous dimensions.
It is worth recalling that the $\Cc_{qrs}$ structures that remain in 
(\ref{gambargen}) are, \cite{12,13},
\begin{eqnarray}
\left. \frac{}{} \Cc_{000} \right|_{d=4} &=&  {\cal I} \otimes {\cal I} 
\otimes {\cal I} \nonumber \\
\left. \frac{}{} \Cc_{220} \right|_{d=4} &=& 
\sigma^{\mu\nu} \otimes \sigma_{\mu\nu} \otimes {\cal I} ~+~
\sigma^{\mu\nu} \otimes {\cal I} \otimes \sigma_{\mu\nu} ~+~
{\cal I} \otimes \sigma^{\mu\nu} \otimes \sigma_{\mu\nu} \nonumber \\
\left. \frac{}{} \Cc_{440} \right|_{d=4} &=& 24 \left[ 
\gamma^5 \otimes \gamma^5 \otimes {\cal I} ~+~
\gamma^5 \otimes {\cal I} \otimes \gamma^5 ~+~
{\cal I} \otimes \gamma^5 \otimes \gamma^5 \right] \nonumber \\
\left. \frac{}{} \Cc_{444} \right|_{d=4} &=& 0 
\end{eqnarray}
in four dimensions.

{\begin{table}[ht]
\begin{center}
\begin{tabular}{|c|c||c|c|c|c|}
\hline
Spin & Chirality & $\Cc_{000}$ & $\Cc_{220}$ & $\Cc_{440}$ & $\Cc_{444}$ \\ 
\hline
$(\half,0)$ & $+$ & 1 & $12$ & $72$ & 0 \\ 
$(\half,0)$ & $-$ & 1 & $12$ & $-24$ & 0 \\ 
$(\threehalves,0)$ & $+$ & 1 & $-12$ & $72$ & 0 \\ 
$(1,\half)$ & $-$ & 1 & $-4$ & $-24$ & 0 \\ 
\hline
\end{tabular}
\caption{Values for the evaluation of the general anomalous dimension for 
various nucleons.}
\label{tabevO}
\end{center}
\end{table}}

The main application of (\ref{gambargen}) is to extract anomalous dimensions 
for four core baryon operators which are central to the evaluation of nucleon 
matrix elements. These were introduced in \cite{11,34} and defined by
\begin{eqnarray}
{\cal O}_+^{(\half,0)} &=& \epsilon^{IJK} \psi_L^I \left( \left( \psi_L^J
\right)^T C \psi_L^K \right) ~~~,~~~ 
{\cal O}_-^{(\half,0)} ~=~ \epsilon^{IJK} \psi_R^I \left( \left( \psi_L^J
\right)^T C \psi_L^K \right) \nonumber \\
{\cal O}_+^{(\threehalves,0)} &=& \epsilon^{IJK} \Deltaslash \psi_L^I 
\Deltaslash \psi_L^J \Deltaslash \psi_L^K ~~~,~~~ 
{\cal O}_-^{(1,\half)} ~=~ \epsilon^{IJK} \Deltaslash \psi_L^I 
\Deltaslash \psi_L^J \Deltaslash \psi_R^K 
\end{eqnarray}
where $\Delta^2$~$=$~$0$. The right and left handed quarks are denoted by 
$\psi_R$~$=$~$\half(1+\gamma^5)\psi$ and $\psi_L$~$=$~$\half(1-\gamma^5)\psi$ 
respectively, the first superscript represents the spin that is not part of
the bilinear part of the operator and the second relates to the bilinear
properties itself, and the subscript sign indicates the chirality \cite{12,20}.
For more details on the discrete and continuous symmetries of these operators 
see the discussion in \cite{20}. To extract the respective anomalous dimensions
merely requires substituting the eigenvalues of the respective $\Cc_{qrs}$ 
structures that were deduced in \cite{12,20,35} and recorded in Table 
\ref{tabevO}. Consequently we have
\begin{eqnarray}
\gamma^{(\half,0)}_+(a) &=& 
4 a
+ \left[
\frac{32}{9} \Nf
- \frac{140}{3}
\right] a^2
+ \left[
160 \Nf
- 32 \zeta_3
- \frac{10784}{9}
- \frac{160}{3} \zeta_3 \Nf
- \frac{112}{27} \Nf^2
\right] a^3
\nonumber \\
&&
+ \left[
528 \zeta_4
+ 848 \zeta_4 \Nf
- \frac{4928575}{162}
- \frac{86600}{27} \zeta_5
- \frac{58972}{27} \zeta_3 \Nf
- \frac{29195}{243} \Nf^2
- \frac{320}{81} \Nf^3
\right. \nonumber \\
&& \left. ~~~~
- \frac{160}{3} \zeta_4 \Nf^2
+ \frac{64}{27} \zeta_3 \Nf^3
+ \frac{320}{27} \zeta_3 \Nf^2
+ \frac{18880}{9} \zeta_5 \Nf
+ \frac{63670}{27} \zeta_3
+ \frac{379232}{81} \Nf
\right] a^4 \nonumber \\
&& +~ O(a^5) \nonumber \\
\gamma^{(\half,0)}_-(a) &=& 
4 a
+ \left[
\frac{32}{9} \Nf
- \frac{100}{3}
\right] a^2
+ \left[
\frac{4264}{27} \Nf
- \frac{10988}{9}
- \frac{160}{3} \zeta_3 \Nf
- \frac{112}{27} \Nf^2
\right] a^3
\nonumber \\
&&
+ \left[
880 \zeta_4 \Nf
- 8800 \zeta_5
- \frac{4227355}{162}
- \frac{66836}{27} \zeta_3 \Nf
- \frac{32179}{243} \Nf^2
- \frac{320}{81} \Nf^3
- \frac{160}{3} \zeta_4 \Nf^2
\right. \nonumber \\
&& \left. ~~~~
+ \frac{64}{27} \zeta_3 \Nf^3
+ \frac{320}{27} \zeta_3 \Nf^2
+ \frac{18400}{9} \zeta_5 \Nf
+ \frac{153818}{27} \zeta_3
+ \frac{361576}{81} \Nf
\right] a^4 ~+~ O(a^5) \nonumber \\
\gamma^{(\threehalves,0)}_+(a) &=& 
-~ 4 a
+ \left[
\frac{40}{9} \Nf
- \frac{328}{3}
\right] a^2
+ \left[
\frac{2008}{9} \Nf
- 2382
- \frac{8}{27} \Nf^2
+ \frac{160}{3} \zeta_3 \Nf
+ \frac{1120}{3} \zeta_3
\right] a^3
\nonumber \\
&&
+ \left[
\frac{40}{81} \Nf^3
- 6160 \zeta_4
- \frac{9495365}{162}
- \frac{36607}{243} \Nf^2
- \frac{26080}{9} \zeta_5 \Nf
- \frac{3200}{27} \zeta_3 \Nf^2
- \frac{1520}{3} \zeta_4 \Nf
\right. \nonumber \\
&& \left. ~~~~
- \frac{64}{27} \zeta_3 \Nf^3
+ \frac{160}{3} \zeta_4 \Nf^2
+ \frac{98720}{27} \zeta_3 \Nf
+ \frac{270644}{27} \zeta_3
+ \frac{293120}{27} \zeta_5
+ \frac{675982}{81} \Nf
\right] a^4 \nonumber \\
&& +~ O(a^5) \nonumber \\
\gamma^{(1,\half)}_-(a) &=& 
-~ \frac{4}{3} a
+ \left[
\frac{112}{27} \Nf
- \frac{236}{3}
\right] a^2
+ \left[
\frac{160}{9} \zeta_3 \Nf
+ \frac{544}{3} \zeta_3
+ \frac{16168}{81} \Nf
- \frac{18496}{9}
- \frac{128}{81} \Nf^2
\right] a^3
\nonumber \\
&&
+ \left[
\frac{565420}{81} \Nf
- 2992 \zeta_4
- 112 \zeta_4 \Nf
- \frac{22115885}{486}
- \frac{36331}{243} \Nf^2
- \frac{27040}{27} \zeta_5 \Nf
\right. \nonumber \\
&& \left. ~~~~
- \frac{6080}{81} \zeta_3 \Nf^2
- \frac{80}{81} \Nf^3
- \frac{64}{81} \zeta_3 \Nf^3
+ \frac{160}{9} \zeta_4 \Nf^2
+ \frac{119804}{81} \zeta_3 \Nf
+ \frac{388640}{243} \zeta_5
\right. \nonumber \\
&& \left. ~~~~
+ \frac{2271070}{243} \zeta_3
\right] a^4 ~+~ O(a^5) 
\end{eqnarray}
where we reproduced the lower loop results of \cite{11,12,13} which represents 
a minor check since \cite{13} did not use {\sc Forcer}. Another check is 
provided by the large $\Nf$ result of \cite{36} where the coefficients of the 
leading order $\Nf$ terms of $\gamma^{(\half,0)}_\pm(a)$ were computed. It is
therefore reassuring to note the coefficient of $\Nf^3$ in the four loop term 
of both these anomalous dimensions is in agreement with the second expression 
of equation (21) in \cite{36}. Numerically we have
\begin{eqnarray}
\gamma^{(\half,0)}_+(a) &=&
4.000000 a
+ (3.555556 \Nf - 46.666667) a^2          
\nonumber \\
&&
+~ ( - 4.148148 \Nf^2 + 95.890298 \Nf - 1236.688043) a^3            
\nonumber \\
&&
+~ ( - 1.101297 \Nf^3 - 163.621338 \Nf^2 + 5149.460288 \Nf - 30343.057304) a^4 ~+~ O(a^5) \nonumber \\
\gamma^{(\half,0)}_-(a) &=& 
4.000000 a
+ (3.555556 \Nf - 33.333333) a^2          
\nonumber \\
&&
+~ ( - 4.148148 \Nf^2 + 93.8162244 \Nf - 1220.888889) a^3           
\nonumber \\
&&
+~ ( - 1.101297 \Nf^3 - 175.901174 \Nf^2 + 4560.706306 \Nf
- 28371.674539) a^4 ~+~ O(a^5) \nonumber \\
\gamma^{(\threehalves,0)}_+(a) &=& 
-~ 4.000000 a
+ (4.444444 \Nf - 109.333333) a^2         
\nonumber \\
&&
+~ ( - 0.296296 \Nf^2 + 287.220813 \Nf - 1933.232089) a^3            
\nonumber \\
&&
+~ ( - 2.355493 \Nf^3 - 235.388188 \Nf^2 + 9187.369682 \Nf 
- 41974.040055) a^4 ~+~ O(a^5) \nonumber \\
\gamma^{(1,\half)}_-(a) &=&
-~ 1.333333 a      
+ (4.148148 \Nf - 78.666667) a^2          
\nonumber \\
&&
+~ ( - 1.580247 \Nf^2 + 220.974839 \Nf - 1837.138126) a^3         
\nonumber \\
&&
+~ ( - 1.937428 \Nf^3 - 220.497455 \Nf^2 + 7598.726041 \Nf - 35851.461429) a^4
\nonumber \\
&& +~ O(a^5)
\end{eqnarray}
in the $\MSbar$ scheme.

\sect{Banks-Zaks fixed point.}

As an application of our baryon operator anomalous dimensions we examine them
at the Banks-Zaks fixed point \cite{26,37}. This is a fixed point in the
situation where the two loop term of the $\beta$-function for a non-abelian
asymptotically free gauge theory is positive. It defines the conformal window. 
From the $SU(3)$ two loop $\beta$-function the window lies in the range 
$8$~$<$~$\Nf$~$\leq$~$16$. The lower boundary is determined from the two loop 
term and there has been a long debate as to whether that boundary remains there
when non-perturbative contributions are included. However a technique was 
introduced in \cite{26} where perturbation theory is carried out in terms of a 
purely dimensionless $\Nf$ dependent parameter that allows one to compute
critical exponents of anomalous dimensions. In our $SU(3)$ case we take
\begin{equation}
\Delta_{\BZss} = 11 ~-~ \frac{2}{3} \Nf 
\label{deldef}
\end{equation}
as our convention for this parameter. The leading order Banks-Zaks critical 
coupling constant is determined by redefining $\Nf$ in the two loop 
$\beta$-function using (\ref{deldef}) and finding the non-trivial solution of 
$\beta(a)$~$=$~$0$ as a function of $\Delta_{\BZss}$. Corrections to the 
leading order critical coupling $a_{\BZss}$ are deduced by including the high 
order terms in the $\beta$-function. As we have determined the baryon 
dimensions to four loops this means we need the five loop $\beta$-function to 
deduce the fourth order term in the associated critical exponents. The reason 
for this is that the one and two loop terms are used to determine the leading 
term of $a_{\BZss}$. Consequently for $SU(3)$ we find
\begin{eqnarray}
a_{\BZss} &=& \frac{\Delta_{\BZss}}{107}
+ \frac{11675}{29401032} \Delta_{\BZss}^2 
+ [ 219204480 \zeta_3 + 145645559 ] \frac{\Delta_{\BZss}^3}{1346449661472} 
\nonumber \\
&&
+~ [ 158627906059008 \zeta_3 - 242692714598400 \zeta_5 + 119816461287557 ]
\frac{\Delta_{\BZss}^4}{6659496939251344896} \nonumber \\
&& +~ O(\Delta_{\BZss}^5) ~.
\end{eqnarray}
Using this value we can deduce the anomalous dimension of the general operator
at the Banks-Zaks fixed point which is
\begin{eqnarray}
\gamma_{\cal O}(a_{\BZss}) &=&
\left[ 
\frac{1}{321} \Delta_{\BZss}
+ \frac{27083}{88203096} \Delta_{\BZss}^2
+ \left[
\frac{352124197}{12118046953248}
- \frac{59531}{2359432818} \zeta_3
\right] \Delta_{\BZss}^3
\right. \nonumber \\
&& \left. ~
+ \left[
\frac{21882082084039}{6659496939251344896}
+ \frac{79450145}{4544267607468} \zeta_5
- \frac{20367037045}{1944946535996304} \zeta_3
\right] \Delta_{\BZss}^4
\right] \Cc_{220} 
\nonumber \\
&& 
+ \left[ 
- \frac{6}{11449} \Delta_{\BZss}^2
+ \left[
\frac{72973}{2359432818}
- \frac{34}{1225043} \zeta_3
\right] \Delta_{\BZss}^3
\right. \nonumber \\
&& \left. ~~~~
+ \left[
\frac{7103828059}{1296631023997536}
- \frac{75332681}{1514755869156} \zeta_3
+ \frac{358930}{3539149227} \zeta_5
\right] \Delta_{\BZss}^4
\right] \Cc_{000} 
\nonumber \\
&& 
+ \left[ 
\frac{1}{412164} \Delta_{\BZss}^2
+ \left[
\frac{71}{33076161} \zeta_3
- \frac{47365}{169879162896}
\right] \Delta_{\BZss}^3
\right. \nonumber \\
&& \left. ~~~~
+ \left[
\frac{6326616091}{93357433727822592}
+ \frac{25858273}{27265605644808} \zeta_3
- \frac{244135}{127409372172} \zeta_5
\right] \Delta_{\BZss}^4
\right] \Cc_{220}^2 
\nonumber \\
&& 
+ \left[ 
- \frac{5}{412164} \Delta_{\BZss}^2
+ \left[
\frac{16525}{56626387632}
- \frac{29}{14700516} \zeta_3
\right] \Delta_{\BZss}^3
\right. \nonumber \\
&& \left. ~~~~
+ \left[
\frac{179045}{84939581448} \zeta_5
- \frac{4311271033}{31119144575940864}
- \frac{37679737}{36354140859744} \zeta_3
\right] \Delta_{\BZss}^4
\right] \Cc_{440} 
\nonumber \\
&& 
+ \left[ 
\left[
\frac{5}{793827864}
- \frac{1}{33076161} \zeta_3
\right] \Delta_{\BZss}^3
\right. \nonumber \\
&& \left. ~~~~
+ \left[
\frac{4205}{254818744344} \zeta_5
+ \frac{164671}{54531211289616} \zeta_3
- \frac{636697}{72708281719488}
\right] \Delta_{\BZss}^4
\right] \Cc_{220}^3 
\nonumber \\
&& 
+ \left[ 
\left[
\frac{25}{176406192} \zeta_3
- \frac{37}{529218576}
\right] \Delta_{\BZss}^3
\right. \nonumber \\
&& \left. ~~~~
+ \left[
\frac{3386513}{436249690316928}
+ \frac{10299631}{145416563438976} \zeta_3
- \frac{16165}{113252775264} \zeta_5
\right] \Delta_{\BZss}^4
\right] \Cc_{220} \Cc_{440} 
\nonumber \\
&& 
+ \left[ 
\left[
\frac{2}{11025387} \zeta_3
- \frac{1}{9800344}
\right] \Delta_{\BZss}^3
\right. \nonumber \\
&& \left. ~~~~
+ \left[
\frac{3564589}{72708281719488}
- \frac{1508161}{18177070429872} \zeta_3
- \frac{48235}{169879162896} \zeta_5
\right] \Delta_{\BZss}^4
\right] \Cc_{444} 
\nonumber \\
&& 
+ \left[ 
\left[
\frac{929}{2038549954752} \zeta_3
- \frac{605}{2038549954752} \zeta_5
- \frac{101}{4077099909504}
\right] \Cc_{220}^4 
\right. \nonumber \\
&& \left. ~~~~
+ \left[
\frac{1015}{1019274977376} \zeta_5
- \frac{2105}{2038549954752} \zeta_3
- \frac{13}{75501850176}
\right] \Cc_{220}^2 \Cc_{440} 
\right. \nonumber \\
&& \left. ~~~~
+ \left[
\frac{895}{254818744344} \zeta_5
- \frac{2009}{509637488688} \zeta_3
- \frac{65}{169879162896}
\right] \Cc_{444} \Cc_{220}
\right. \nonumber \\
&& \left. ~~~~
+ \left[
\frac{865}{4077099909504}
- \frac{1025}{2038549954752} \zeta_5
- \frac{235}{2038549954752} \zeta_3
\right] \Cc_{440}^2
\right] \Delta_{\BZss}^4 \nonumber \\
&& +~ O(\Delta_{\BZss}^5)
\end{eqnarray}
or
\begin{eqnarray}
\gamma_{\cal O}(a_{\BZss}) &=&
3.115265 \times 10^{-3} \, \Cc_{220} \Delta_{\BZss} 
\nonumber \\
&&
+ \left[ - 5.240632 \times 10^{-4} \, \Cc_{000} 
+ 2.426219 \times 10^{-6} \, \Cc_{220}^2 
+ 3.070527 \times 10^{-4} \, \Cc_{220} 
\right. \nonumber \\
&& \left. ~~~~
- 1.213109 \times 10^{-5} \, \Cc_{440} \right] \Delta_{\BZss}^2 
\nonumber \\
&&
+ \left[ - 2.433845 \times 10^{-6} \, \Cc_{000} 
- 3.004350 \times 10^{-8} \, \Cc_{220}^3 
+ 2.301473 \times 10^{-6} \, \Cc_{220}^2 
\right. \nonumber \\
&& \left. ~~~~
+ 1.004392 \times 10^{-7} \, \Cc_{220} \Cc_{440} 
- 1.271340 \times 10^{-6} \, \Cc_{220} 
- 2.079496 \times 10^{-6} \, \Cc_{440} 
\right. \nonumber \\
&& \left. ~~~~
+ 1.160153 \times 10^{-7} \, \Cc_{444} \right] \Delta_{\BZss}^3 
\nonumber \\
&&
+ \left[ 5.085945 \times 10^{-5} \, \Cc_{000} 
+ 2.152852 \times 10^{-10} \, \Cc_{220}^4 
+ 1.198435 \times 10^{-8} \, \Cc_{220}^3 
\right. \nonumber \\
&& \left. ~~~~
- 3.808425 \times 10^{-10} \, \Cc_{220}^2 \Cc_{440} 
- 7.791257 \times 10^{-7} \, \Cc_{220}^2 
- 5.510202 \times 10^{-8} \, \Cc_{220} \Cc_{440} 
\right. \nonumber \\
&& \left. ~~~~
- 1.479153 \times 10^{-9} \, \Cc_{444} \Cc_{220}
+ 8.827407 \times 10^{-6} \, \Cc_{220} 
- 4.477861 \times 10^{-10} \, \Cc_{440}^2 
\right. \nonumber \\
&& \left. ~~~~
+ 8.013217 \times 10^{-7} \, \Cc_{440} 
- 3.451317 \times 10^{-7} \, \Cc_{444} 
\right] \Delta_{\BZss}^4 ~+~ O(\Delta_{\BZss}^5) ~.
\end{eqnarray}
numerically. Consequently the anomalous dimensions of the four core baryon
operators are
\begin{eqnarray}
\gamma^{(\half,0)}_+(a_{\BZss}) &=& 
\frac{4}{107} \Delta_{\BZss}
+ \frac{19379}{7350258} \Delta_{\BZss}^2
+ \left[
\frac{314021069}{1009837246104}
- \frac{12224}{131079601} \zeta_3
\right] \Delta_{\BZss}^3
\nonumber \\
&&
+ \left[
\frac{19461603055087}{554958078270945408}
- \frac{1981080112}{40519719499923} \zeta_3
+ \frac{35529320}{378688967289} \zeta_5
\right] \Delta_{\BZss}^4 \nonumber \\
&& +~ O(\Delta_{\BZss}^5) \nonumber \\
\gamma^{(\half,0)}_-(a_{\BZss}) &=& 
\frac{4}{107} \Delta_{\BZss}
+ \frac{9313}{2450086} \Delta_{\BZss}^2
+ \left[
\frac{40784885}{112204138456}
- \frac{8800}{131079601} \zeta_3
\right] \Delta_{\BZss}^3
\nonumber \\
&&
+ \left[
\frac{22658041606303}{554958078270945408}
- \frac{218534068}{13506573166641} \zeta_3
+ \frac{1870000}{42076551921} \zeta_5
\right] \Delta_{\BZss}^4 \nonumber \\
&& +~ O(\Delta_{\BZss}^5) \nonumber \\
\gamma^{(\threehalves,0)}_+(a_{\BZss}) &=& 
-~ \frac{4}{107} \Delta_{\BZss}
- \frac{34787}{7350258} \Delta_{\BZss}^2
+ \left[
\frac{146240}{393238803} \zeta_3
- \frac{32245429}{112204138456}
\right] \Delta_{\BZss}^3
\nonumber \\
&&
+ \left[
\frac{2820290956}{40519719499923} \zeta_3
- \frac{51566480}{378688967289} \zeta_5
- \frac{14951711813983}{554958078270945408}
\right] \Delta_{\BZss}^4 \nonumber \\
&& +~ O(\Delta_{\BZss}^5) \nonumber \\
\gamma^{(1,\half)}_-(a_{\BZss}) &=& 
-~ \frac{4}{321} \Delta_{\BZss}
- \frac{31363}{22050774} \Delta_{\BZss}^2
+ \left[
\frac{22336}{131079601} \zeta_3
- \frac{314714429}{3029511738312}
\right] \Delta_{\BZss}^3
\nonumber \\
&&
+ \left[
\frac{14316089632}{364677475499307} \zeta_3
- \frac{222460400}{3408200705601} \zeta_5
- \frac{109005718637}{61662008696771712}
\right] \Delta_{\BZss}^4 \nonumber \\
&& +~ O(\Delta_{\BZss}^5) ~.
\end{eqnarray}
Expressing these numerically leads to
\begin{eqnarray}
\gamma^{(\half,0)}_+(a_{\BZss}) &=& 
3.738318\times 10^{-2} \, \Delta_{\BZss}
+ 2.636506 \times 10^{-3} \, \Delta_{\BZss}^2 
+ 1.988627 \times 10^{-4} \, \Delta_{\BZss}^3 
\nonumber \\
&&
+~ 7.358447 \times 10^{-5} \, \Delta_{\BZss}^4 ~+~ O(\Delta_{\BZss}^5)
\nonumber \\
\gamma^{(\half,0)}_-(a_{\BZss}) &=& 
3.738318 \times 10^{-2} \, \Delta_{\BZss}
+ 3.801091 \times 10^{-3} \, \Delta_{\BZss}^2 
+ 2.827884 \times 10^{-4} \, \Delta_{\BZss}^3 
\nonumber \\
&&
+~ 6.746328 \times 10^{-5} \, \Delta_{\BZss}^4  ~+~ O(\Delta_{\BZss}^5)
\nonumber \\
\gamma^{(\threehalves,0)}_+(a_{\BZss}) &=& 
-~ 3.738318 \times 10^{-2} \, \Delta_{\BZss}
- 4.732759 \times 10^{-3} \, \Delta_{\BZss}^2 
+ 1.596463 \times 10^{-4} \, \Delta_{\BZss}^3 
\nonumber \\
&&
-~ 8.447494 \times 10^{-5} \, \Delta_{\BZss}^4  ~+~ O(\Delta_{\BZss}^5)
\nonumber \\
\gamma^{(1,\half)}_-(a_{\BZss}) &=&
-~ 1.246106\times 10^{-2} \, \Delta_{\BZss}
- 1.422308 \times 10^{-3} \, \Delta_{\BZss}^2 
+ 1.009479 \times 10^{-4} \, \Delta_{\BZss}^3 
\nonumber \\
&&
-~ 2.226127 \times 10^{-5} \, \Delta_{\BZss}^4 ~+~ O(\Delta_{\BZss}^5)
\end{eqnarray}
from which it can be observed that the coefficients of each operator dimension
diminish by roughly an order of magnitude with respect to this expansion 
parameter. Moreover as $\Delta_{\BZss}$ ranges from $1/3$ to $17/3$ for 
$\Nf$~$=$~$16$ down to $\Nf$~$=$~$8$ across the conformal window it may appear 
that the series convergence should be reasonable. In order to gauge this we 
have plotted the anomalous dimensions for each of the operators in Figure 
\ref{anomdimfig} for the leading order (LO) to the fourth order term which is 
next-to-next-to-next-to leading order (NNNLO). We note the leading order is 
linear in $\Nf$ and so has a limited contribution to the convergence 
discussion. 

{\begin{figure}[H]
{\begin{center}
\includegraphics[width=7.80cm,height=7cm]{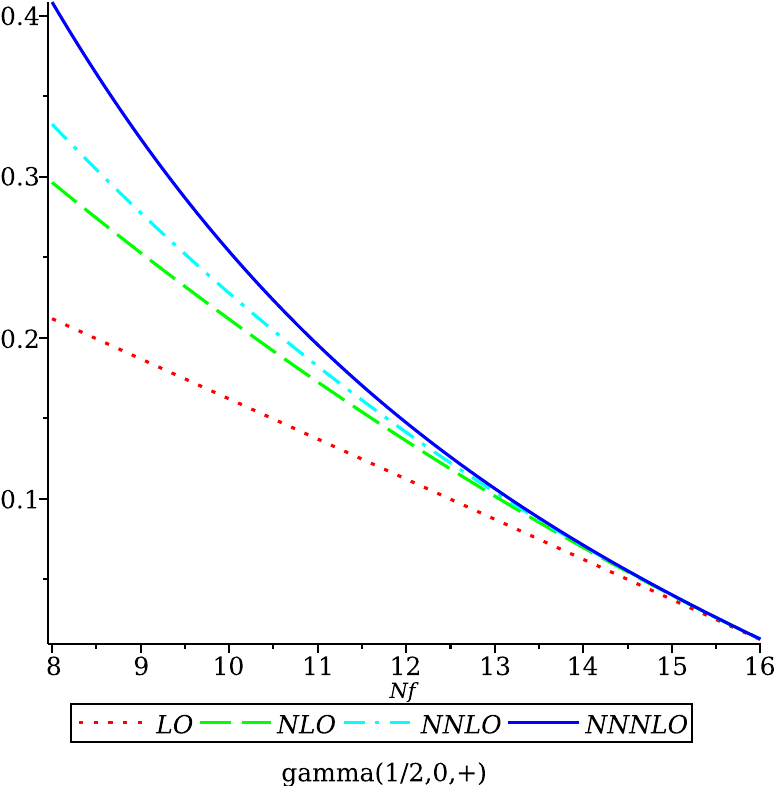}
\quad
\quad
\includegraphics[width=7.80cm,height=7cm]{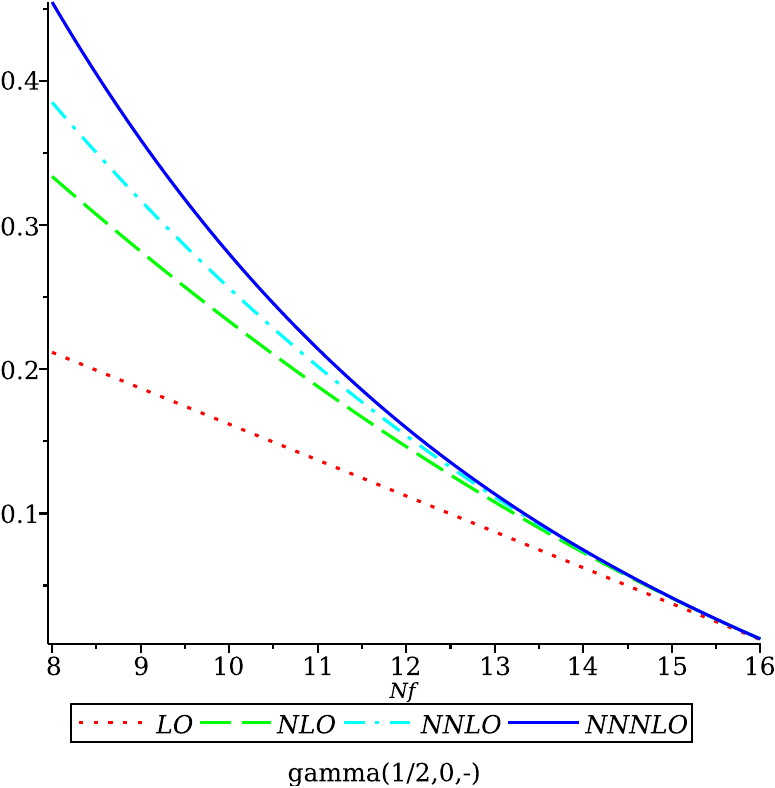}

\vspace{1cm}
\includegraphics[width=7.80cm,height=7cm]{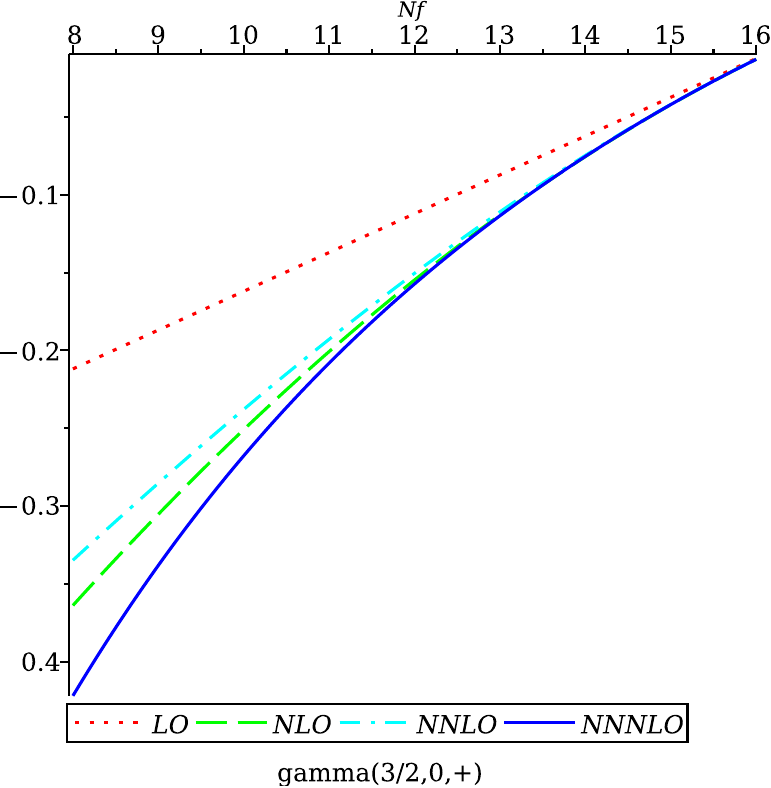}
\quad
\quad
\includegraphics[width=7.80cm,height=7cm]{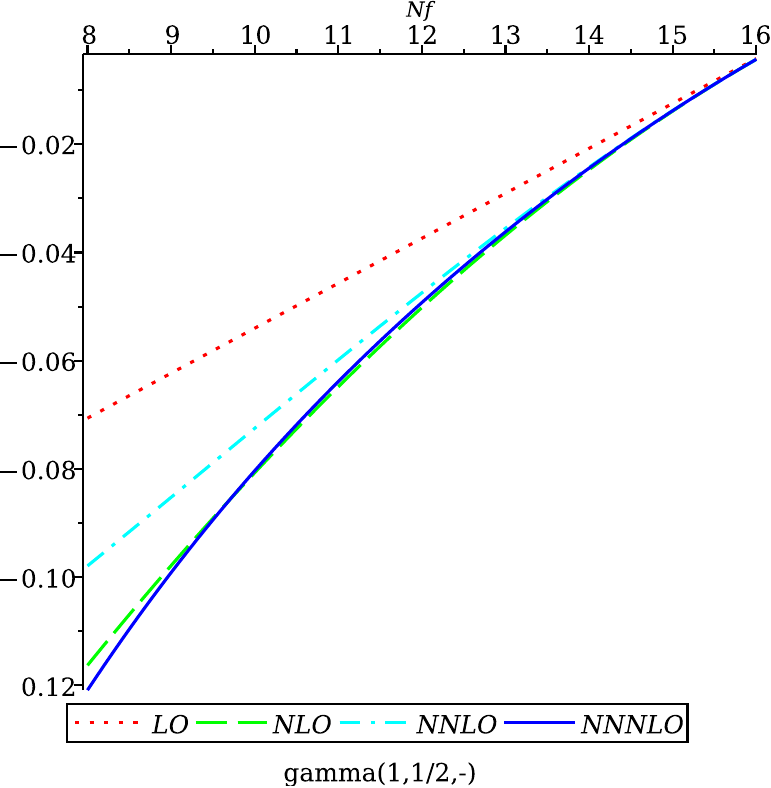}
\caption{Plots of the leading order (LO), next-to-leading order (NLO),
next-to-next-to-leading order (NNLO), next-to-next-to-next-to-leading order 
(NNNLO), anomalous dimension of the four baryon operators at the Banks-Zaks 
fixed point for $8$~$\leq$~$\Nf$~$\leq$~$16$.}
\label{anomdimfig}
\end{center}}
\end{figure}}

What is evident from Figure \ref{anomdimfig} is that for each of the four
operators perturbation theory appears to be reliable down to around 
$\Nf$~$=$~$12$. While if anything it may seem that the behaviour of the two 
spin-$\half$ cases are both less accurate at this value than either the 
spin-$1$ or spin-$\threehalves$ operator dimensions this is not the case. 
Examining the lower plots of Figure \ref{anomdimfig} there is a wide
fluctuation below $\Nf$~$=$~$12$ since the second and fourth order lines are
particularly close especially in the spin-$1$ situation. This contrasts with
the spin-$\half$ plots where there is no fluctuation of the order. In order to
explore the convergence of each series we have undertaken a Pad\'{e} 
approximant study on the fourth order expression which will provide three 
examples. This is because the leading order is $O(\Delta_{\BZss})$. So we will 
construct the $[1,1]$, $[1,2]$ and $[2,2]$ approximants from the second, third 
and fourth order approximations. The numerical values of the resulting
exponents are provided in Table \ref{tabpadbary}. It is clear that the 
convergence for the two spin-$\half$ operators is improved for 
$\Nf$~$\leq$~$12$ and particularly so for the positive chirality case even for 
$\Nf$~$=$~$8$. Though in this specific case the situation has to be qualified 
by noting that this value of $\Nf$ may be outside the actual conformal window.
As yet there has been no definitive determination as to what that value is. To 
understand these comments further and to illustrate the improvement provided by
the Pad\'{e} approximant we have plotted the graphs of the three approximants 
of each of the two spin-$\half$ operators in Figure \ref{figpadbary}. The 
distinctly more accurate convergence in the positive chirality case is 
apparent. For the remaining two operators the convergence of both their 
approximants appear not to have improved below $\Nf$~$=$~$14$ which is why we 
have not provided any illustrations of those approximants. In essence the 
order-by-order discrepancies in Table \ref{tabpadbary} is a reflection of this.
Finally we note that the respective unitarity bounds of each of the operators, 
which were derived in \cite{20} and checked to third order, remain unviolated 
at fourth order in the conformal window range.

{\begin{table}[ht]
\begin{center}
\begin{tabular}{|c|c||c|c|c|}
\hline
Exponent & $\Nf$ & $P_{[1,1]}$ & $P_{[1,2]}$ & $P_{[2,2]}$ \\ 
\hline
\rule{0pt}{20pt}
$\gamma^{(\half,0)}_+(a_{\BZss})$ 
& $8$ &
$0.352858$ &
$0.359503$ &
$0.352594$ \\
& $10$ &
$0.233291$ &
$0.235492$ &
$0.233174$ \\
& $12$ &
$0.142246$ &
$0.142809$ &
$0.142201$ \\
& $14$ &
$0.070604$ & 
$0.070681$ & 
$0.070593$ \\ 
& $16$ & 
$0.012761$ & 
$0.012761$ & 
$0.012760$ \\
\hline
\rule{0pt}{20pt}
$\gamma^{(\half,0)}_-(a_{\BZss})$ 
& $8$ & 
$0.499833$ & 
$0.413023$ & 
$0.472906$ \\ 
& $10$ & 
$0.289590$ & 
$0.264920$ & 
$0.281032$ \\ 
& $12$ & 
$0.161375$ & 
$0.155779$ & 
$0.159019$ \\ 
& $14$ & 
$0.075018$ & 
$0.074329$ & 
$0.074632$ \\ 
& $16$ & 
$0.012898$ & 
$0.012894$ & 
$0.012895$ \\
\hline
\rule{0pt}{20pt}
$\gamma^{(\threehalves,0)}_+(a_{\BZss})$ 
& $8$ & 
$ - 0.749623$ & 
$ - 0.226711$ & 
$ - 0.389838$ \\ 
& $10$ & 
$ - 0.358874$ & 
$ - 0.194575$ & 
$ - 0.257553$ \\ 
& $12$ & 
$ - 0.180829$ & 
$ - 0.139684$ & 
$ - 0.155322$ \\ 
& $14$ & 
$ - 0.078968$ & 
$ - 0.073701$ & 
$ - 0.075236$ \\ 
& $16$ & 
$ - 0.013010$ & 
$ - 0.012980$ & 
$ - 0.012982$ \\
\hline
\rule{0pt}{20pt}
$\gamma^{(1,\half)}_-(a_{\BZss})$ 
& $8$ & 
$ - 0.199920$ & 
$ - 0.068444$ & 
$ - 0.110503$ \\ 
& $10$ & 
$ - 0.106844$ & 
$ - 0.059855$ & 
$ - 0.077264$ \\  
& $12$ & 
$ - 0.056850$ & 
$ - 0.044097$ & 
$ - 0.048697$ \\ 
& $14$ & 
$ - 0.025647$ & 
$ - 0.023914$ & 
$ - 0.024388$ \\
& $16$ & 
$ - 0.004318$ & 
$ - 0.004307$ & 
$ - 0.004308$ \\
\hline
\end{tabular}
\caption{Pad\'{e} estimates for the anomalous dimension of the four baryon 
operators at the Banks-Zaks fixed point for $8$~$\leq$~$\Nf$~$\leq$~$16$.}
\label{tabpadbary}
\end{center}
\end{table}}

{\begin{figure}[H]
{\begin{center}
\includegraphics[width=7.80cm,height=7cm]{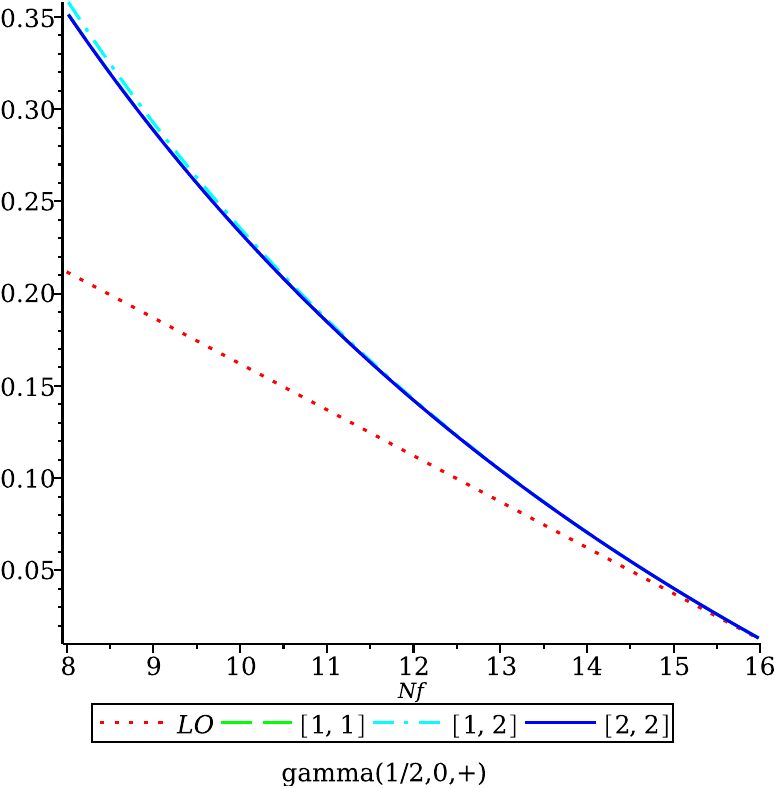}
\quad
\quad
\includegraphics[width=7.80cm,height=7cm]{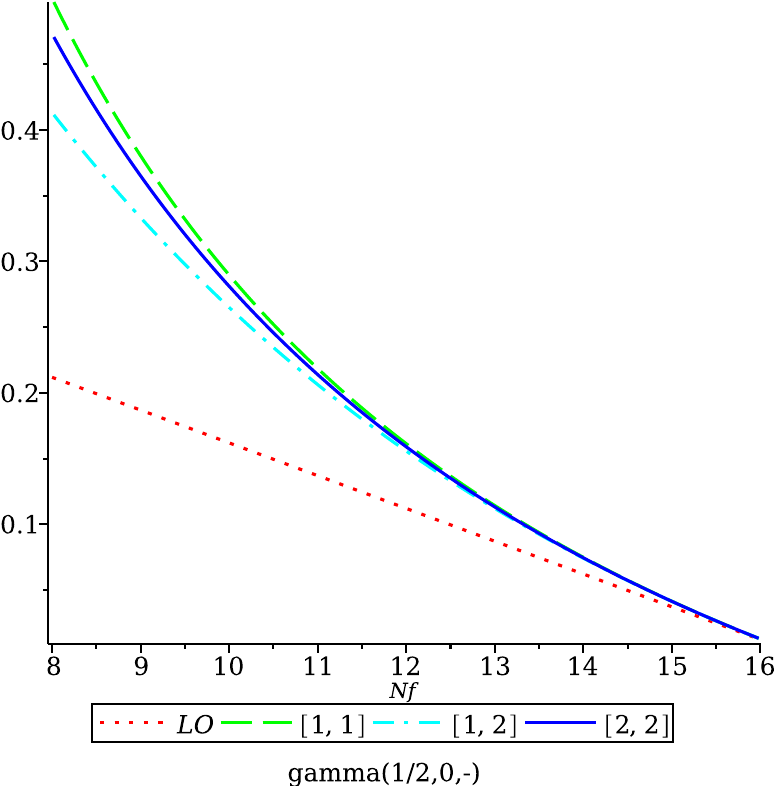}
\caption{Plots of the Pad\'{e} estimates for the anomalous dimension of the two
spin-$\half$ baryon operators at the Banks-Zaks fixed point for 
$8$~$\leq$~$\Nf$~$\leq$~$16$.}
\label{figpadbary}
\end{center}}
\end{figure}}

\sect{Discussion.}

We have completed the four loop $\MSbar$ scheme renormalization of a general
$3$-quark operator using the Manashov and Kr\"{a}nkl approach thereby extending
the two and three loop results of \cite{12,13}. In carrying out the 
renormalization with a general and flexible approach the anomalous dimensions 
of several baryonic bound states have been deduced as a by-product. Since such 
baryon operators have been of interest in exploring ideas for extensions of the
Standard Model, \cite{19,20,21,22,23,24,25}, by way of an application we 
examined the behaviour of the eigen-operator anomalous dimensions in the 
conformal window of QCD using Banks-Zaks perturbation theory, \cite{26}. 
Ordinarily from similar studies, such as \cite{38} for example, perturbation 
theory is reliable from the top of the window down to around about 
$\Nf$~$=$~$12$. By this we mean that the leading, next-to-leading, 
next-to-next-to-leading and next-to-next-to-next-to-leading approximations 
produce critical exponent estimates that are in close agreement. Below this 
value of $\Nf$, where the Banks-Zaks fixed point starts to drift beyond the 
perturbative approximation, the estimates begin to diverge. What was 
interesting with several of the baryon operator critical exponents was that 
Pad\'{e} approximants appeared to improve the convergence down to the 
neighbourhood of what is perceived to be the lower conformal window boundary. 
However this was very much dependent on the actual operator itself rather than 
being a global property. What would be interesting to ascertain is whether the 
Pad\'{e} estimates of the ${\cal O}_+^{(\half,0)}$ and ${\cal O}_-^{(\half,0)}$
dimensions could be measured using lattice field theory for $\Nf$~$=$~$10$ for
instance. The reason being that from lattice field theory studies this $\Nf$ 
value is believed to lie inside the conformal window. If so it would give 
momentum to following a similar approach for other useful operators in 
conformal window analyses of Standard Model extensions. Separately another 
thread that needs to be followed is to extend the evaluation of baryon operator
dimensions to the determination of their higher moments.

\vspace{1cm}
\noindent
{\bf Data Availability Statement.} The data that support the findings of this
article are openly available \cite{39}.

\vspace{1cm}
\noindent
{\bf Acknowledgement.} The author thanks the DESY Zeuthen theory group, where 
part of this work was carried out, for their hospitality and Kolleg Mathematik
Physik Berlin for a Visiting Scholarship. We thank L. Vecchi for 
correspondence. The research was carried out with the support of the STFC 
Consolidated Grant ST/X000699/1 and was undertaken on {\sc Barkla}, part of the
High Performance Computing facilities at the University of Liverpool, UK. For 
the purpose of open access, the author has applied a Creative Commons 
Attribution (CC-BY) licence to any Author Accepted Manuscript version arising.

\appendix

\sect{Products of $\Cc_{qrs}$.}

In this appendix we record the products of $\Cc_{qrs}$ which were required to
convert the operator renormalization constant to its anomalous dimension
in (\ref{ztogamma}). We have 
\begin{eqnarray}
\Cc_{220} \Cc_{220} &=& 6 d (d - 1) \Cc_{000} ~+~ 4 (d - 3) \Cc_{220} ~+~ 
\Cc_{440} ~+~ 2 \Cc_{422} 
\nonumber \\
\Cc_{440} \Cc_{220} &=& \Cc_{220} \Cc_{440} \nonumber \\
&=& 12 (d - 2) (d - 3) \Cc_{220} ~+~ 8 (d - 4) \Cc_{440} ~-~ 24 \Cc_{422} ~+~ 
\Cc_{660} ~+~ \Cc_{642} 
\nonumber \\
\Cc_{220} \Cc_{422} &=& \Cc_{422} \Cc_{220} \nonumber \\
&=& 4 (d - 2) (d - 3) \Cc_{220} ~-~ 4 \Cc_{440} ~+~ 4 (2 d - 7) \Cc_{422} ~+~
3 \Cc_{444} ~+~ \Cc_{642} 
\nonumber \\
\Cc_{220} \Cc_{660} &=& \Cc_{660} \Cc_{220} \nonumber \\
&=& 30 (d - 4) (d - 5) \Cc_{440} ~+~ 12 (d - 6) \Cc_{660} ~-~ 30 \Cc_{642} ~+~
\Cc_{880} ~+~ \Cc_{862} 
\nonumber \\
\Cc_{220} \Cc_{642} &=& \Cc_{642} \Cc_{220} \nonumber \\
&=& 4 (d - 4) (d - 5) \Cc_{440} ~+~ 24 (d - 4) (d - 5) \Cc_{422} ~-~
72 \Cc_{444} ~-~ 4 \Cc_{660} 
\nonumber \\
&&
+~ 2 (6 d - 35) \Cc_{642} ~+~ \Cc_{862} ~+~ 2 \Cc_{844} ~+~ 2 \Cc_{664} 
\nonumber \\
\Cc_{220} \Cc_{444} &=& \Cc_{444} \Cc_{220} \nonumber \\
&=& 2 (d - 4) (d - 5) \Cc_{422} ~+~ 12 (d - 4) \Cc_{444} ~-~ 2 \Cc_{642} ~+~ 
\Cc_{664} 
\nonumber \\
\Cc_{440} \Cc_{440}
&=& 72 d (d - 1) (d - 2) (d - 3) \Cc_{000} ~+~ 
96 (d - 2) (d - 3) (d - 4) \Cc_{220} 
\nonumber \\
&&
+~ 24 (3 d^2 - 27 d + 62) \Cc_{440} ~-~ 432 \Cc_{444} ~+~ 
16 (d - 6) \Cc_{660} ~+~ \Cc_{880} ~+~ 2 \Cc_{844} 
\nonumber \\
\Cc_{440} \Cc_{422} &=& \Cc_{422} \Cc_{440} \nonumber \\
&=& - 48 (d - 2) (d - 3) \Cc_{220} ~+~ 24 (d^2 - 13 d + 37) \Cc_{422} ~+~ 
144 \Cc_{444} ~+~ 8 (d - 9) \Cc_{642} 
\nonumber \\
&&
+~ \Cc_{862} ~+~ \Cc_{664} 
\nonumber \\
\Cc_{422} \Cc_{422}
&=& 12 d (d - 1) (d - 2) (d - 3) \Cc_{000} ~+~ 
8 (2 d - 7) (d - 2) (d - 3) \Cc_{220} 
\nonumber \\
&&
+~ 4 (d^2 - 13 d + 37) \Cc_{440} ~+~ 4 (5 d - 19) (d - 2) \Cc_{422} ~+~ 
12 (2 d - 17) \Cc_{444} 
\nonumber \\
&&
-~ 4 \Cc_{660} ~+ 4 (d - 3) \Cc_{642} ~+~ \Cc_{844} ~+~ 2 \Cc_{664} 
\label{Ccprod}
\end{eqnarray}
in addition to the trivial products
\begin{equation}
\Cc_{000} \Cc_{qrs} ~=~ \Cc_{qrs} \Cc_{000} ~=~ \Cc_{qrs}
\end{equation}
for arbitrary $q$, $r$ and $s$. We note the first equation of (\ref{Ccprod}) is 
equivalent to (\ref{gaml2reln}). Also the order of the products of different 
$\Cc_{qrs}$ produces the same expression.

\end{document}